\documentclass[10pt,a4paper]{article}
\usepackage{amssymb}
\newtheorem{prop}{Proposition}
\newtheorem{ex}{Example}
\newtheorem{rk}{Remark}
\newtheorem{thm}{Theorem}
\newtheorem{defn}{Definition}

\newtheorem{lem}{Lemma}

\usepackage{amssymb}
\usepackage{color}
\usepackage{colordvi}
\usepackage{amsmath}

\def\dbar{d{\hskip-1pt\bar{}}\hskip1pt}
\def\cutoffint{-\hskip -10pt\int}
\def \C{{\! \rm \
I \!\!\!C}}

\def \R {{\! \rm \ I \!R}}

\def \N {{\! \rm \ I \!N}}

\def \Z {{\! \rm Z\! \!Z}}
 \def\cutoffint{-\hskip -10pt\int}

\def\otherterm#1{{\it#1}}
\def \e {{\epsilon}}

\def \Ci {{C^\infty}}

\def \Cl {{C\ell}}

\def \endsquare{ $\sqcup \!\!\!\! \sqcap$ }

\def\Ci{C^\infty}

\def\dd{\partial}

\def\Di{D\kern -.65em /}
\def\Dii{D\kern -.45em /}
\def\di{{\dd}\kern -.55em /}
\def\dii{{\dd}\kern -.40em /}

\def\noi{\noindent}

\def\to{\rightarrow}

\def\Dd{{\mathcal D}}

\def\Nn{{\mathcal N}}

\def\Rr{{\mathcal R}}

\def\={\cong}
\def\>{\supset}
\def\<{\subset}

\def\12{\frac{1}{2}}
\def\2{\Dd}
\def\3{\Nn}
\def\4{\Rr}
\def\6{\cup}
\def\8{\otimes}
\def\0{^{\circ}}
\def\){\hfill{\ \qed}\enddemo}

\def\R{\mathbb{R}}
\def\C{\mathbb{C}}

\def\e{\varepsilon}

\def\N{\NN}

\def\Si{\Sigma}

\def\Z{\ZZ}

\def\Cl{\mbox{\rm C$\ell$}}

\def\Dd{{\mathcal D}}

\def\Nn{{\mathcal N}}

\def\Rr{{\mathcal R}}

\def\Si{S\kern -.65em /}

\def\dbar{d{\hskip-1pt\bar{}}\hskip1pt}

\def \C{{\! \rm \ I \!\!\!C}}
\def \R {{\! \rm \ I \!R}}
\def \N {{\! \rm \ I \!N}}

\def \Z {{\! \rm Z\! \!Z}}

\def\cutoffint{-\hskip -12pt\int}

\def\otherterm#1{{\it#1}}
\def \e {{\epsilon}}

\def \Ci {{C^\infty}}

\def \Cl {{C\ell}}

\begin{document}

\title{\bf The  multiplicative anomaly for determinants revisited; locality.}
\author{ Marie-Fran\c coise OUEDRAOGO, Sylvie
PAYCHA  } \maketitle
\section*{Abstract}
 Observing that  the logarithm of a product of two elliptic operators differs from the sum of the
 logarithms by a finite sum of operator brackets, we infer that
  regularised traces of this difference are local as  finite sums of
 noncommutative residues.  From  an explicit local formula for such regularised
 traces,  we derive  an explicit
local formula for the multiplicative anomaly of
$\zeta$-determinants which sheds  light on its locality and yields back previously known results.
\section*{Acknowledgements} We would like to thank Catherine Ducourtioux for
her very useful comments  and for pointing
out to relevant references. We are also grateful to the referee for very
helpful comments on a previous version of this paper.
 \section*{Introduction}
\indent The determinant on the linear group $Gl(\R^n)$  reads
$${\rm det} A= e^{{\rm tr} \left(\log A\right)}$$
where tr is the matrix trace. It is independent  of the choice of
spectral cut used to define the logarithm and is multiplicative as a
result of the Campbell-Hausdorff formula and the cyclicity of the
trace, namely:
$${\rm det} (AB)= e^{{\rm tr} \left(\log AB\right)}= e^{{\rm tr} \left(\log A+\log
B\right)}= {\rm det} A\, {\rm det} B.$$
\indent  In contrast,  the $\zeta$-determinant $${\rm det}_\zeta(A)=e^{-\zeta_A^\prime(0)},$$
defined for an admissible elliptic classical pseudodifferential operator $A$ (with
appropriate spectral cut) acting on sections of a vector bundle $E$ over    a
closed $n$-dimensional 
manifold  $M$ via the zeta function $\zeta_A(s)$   associated with $A$, which corresponds to the unique
meromorphic extension  of the map $s\mapsto {\rm
  Tr}(A^{-s})$ given by the $L^2$-trace of $A^{-s}$ defined on the domain of
holomorphicity  ${\rm Re}(as)>n$ where $a$ is the order of $A$, is not
multiplicative. It 
presents a multiplicative anomaly $${\cal M}_\zeta(A, B)= \frac{{\rm det}_\zeta
  (AB)}{{\rm det}_\zeta(A)\, {\rm det}_\zeta(B)}$$ studied independently by K. Okikiolu in
\cite{O2} and  by M.
Kontsevich and S. Vishik in \cite{KV}.\\ \indent
The  multiplicative anomaly of
$\zeta$-determinants was expressed in terms of
noncommutative residues of classical pseudodifferential operators in the
following situations:
\begin{itemize}
\item by Wodzicki
  \cite{W1} (see also \cite{K} for a review)  for positive definite
commuting elliptic differential operators,
\item by Friedlander \cite{Fr}
for positive definite elliptic pseudodifferential operators,
\item by
Okikiolu  \cite{O2} for operators with scalar leading symbols,
\item  by  Kontsevich and Vishik \cite{KV}
 for operators sufficiently close to self-adjoint
positive pseudodifferential operators.
\item The multiplicative anomaly was further studied by
Ducourtioux  \cite{D} in the context of weighted determinants   also  discussed in
this paper.
\end{itemize}
\indent
The noncommutative residue res (see formula (\ref{eq:resA})) originally
introduced by Guillemin \cite{G} and Wodzicki \cite{W1}, which
defines a trace on the algebra  $\Cl(M, E)$ of classical
pseudodifferential operators acting on smooth sections of the vector bundle
$E$, is local in so far as it is the integral over $M$ of a local residue
${\rm res}_x(A)$ which only depends on a
finite number of homogeneous components of the symbol of the
operator $A$. Consequently, the multiplicative anomaly is local. \\ \indent
Locality of the multiplicative
  anomaly   for $\zeta$-determinants  relates to the locality  of
regularised traces\footnote{Regularised traces   are linear extensions to the algebra
$\Cl(M, E)$ of the ordinary $L^2$
-trace on smoothing operators, which are non tracial since the $L^2$-trace
does not  extend to a trace on the
 algebra $\Cl(M, E)$.} of the difference
$$L(A,B):= \log(AB)-\log A-\log B,$$  on which we focus in  this paper,
investigating their local feature which follows from the vanishing of the
residue of $L(A,B)$.   \\  To see these links, one first observes that regularised traces of $L(A,B)$ correspond to the
 multiplicative anomaly of another type of regularised determinants, namely
 weighted determinants (see \cite{D})
$${\rm det}^Q(A)= e^{{\rm tr}^Q(\log A)},$$
defined via a regularised trace ${\rm tr}^Q$  (see Definition
\ref{defn:weightedtrace}) which uses the regulator $Q$,  called a 
weight \footnote{ A weight is any admissible  elliptic  operator in $\Cl(M, E)$ with positive
order.}.  They differ from $\zeta$-determinants by a local expression
 involving the Wodzicki residue,
 as can  explicitely be seen from the relation (see \cite{D}  Proposition III.1.7):
 $$\frac{{\rm det}_\zeta (A)}{{\rm Det}^Q(A)}= e^{-\frac{1}{2a} {\rm
   res}\left[\left(\log A- \frac{a}{q} \log Q\right)^2\right]},$$
where $a$ is the order of $A$, $q$ the order of $Q$. Consequently, the
multiplicative anomaly for zeta determinants differs from the
multiplicative anomaly for weighted
determinants  by a local expression so that $\log 
  {\cal M}_\zeta(A,B)-{\rm tr}^Q\left(L(A,B)\right)$  is local. 
\\
\indent On the other hand, one infers the locality of regularised traces ${\rm
  tr}^Q(L(A,B))$ of
$L(A,B)$ from 
 the vanishing of the noncommutative residue of $L(A,B)$ (see (\ref{eq:resLvanishes})), a property shown in \cite{Sc} which
 implies the
  multiplicativity of the residue determinant.   Indeed, since  all   traces on the algebra of
  classical pseudodifferential operators on a closed manifold of dimension
  larger than one
 are proportional to the noncommutative residue \cite{W1}, it follows that   $L(A,B)$ is a finite sum
of commutators of classical pseudodifferential  operators. Combining this
with the expression of regularised traces of brackets in terms on the
noncommutative residue (see (\ref{eq:coboundary})), yields the locality of
regularised traces ${\rm
  tr}^Q(L(A,B))$  as 
finite sums of noncommutative residues. \\ \indent
Explicitely, in Theorem \ref{thm:weightedTraceLAB} we show that for  two
admissible elliptic  operators $A, B$    with positive orders
 $a$ and  $b$, such that the product $AB$ is also admissible, there is an operator $
    W(\tau)(A,B):={\frac{d}{dt}}_{\vert {t=0}} L(A^t, A^\tau B)$  depending
    continuously on $\tau$ such that (see (\ref{eq:trQL}))
\begin{equation}\label{eq:introtraceQL}
{\rm tr}^Q(L(A,B))= \int_0^1{\rm
  res}\left(W(\tau)(A,B)\left(\frac{\log(A^\tau B)}{a\tau+b}
-\frac{\log Q}{q}\right)\right)d\tau.
\end{equation}
 \\ \indent
The multiplicative anomaly for weighted determinants 
derived in   Proposition \ref{prop:weightedmultanom}  follows in a
straightforward manner. From (\ref{eq:introtraceQL}),   in Theorem
\ref{thm:zetamultanom} we then derive an explicit local formula for  the multiplicative anomaly for operators $A$ and $B$ with positive orders $a$
and $b$  (see equation (\ref{eq:resmultiplicativeanomaly})):
 \begin{eqnarray} \label{eq:intromultanomaly}
\log {\cal M}_\zeta(A, B)&=& \int_0^1{\rm
  res}\left( W(\tau)(A,B)\,\left(\frac{\log(A^\tau B)}{a\tau+b}
-\frac{\log B}{b}\right)\right) d\tau\nonumber \nonumber\\ 
&+ & {\rm res} \left(
\frac{L(A, B)\,  \log B  }{b} - \frac{\log^2 A\, B}{2(a+b)}-\frac{\log^2A}{2a}- \frac{\log^2B}{2b} \right)
\end{eqnarray}
and similarly with the roles of $A$ and $B$ interchanged. When
the operators $A$ and $B$ commute, $L(A,B)$ vanishes and formula (\ref{eq:intromultanomaly})  yields back Wodzicki's formula:
$$\log {\cal M}_\zeta(A, B)=-
{\rm res} \left(\frac{\log^2 A\, B}{2(a+b)}+\frac{\log^2A}{2a}+ \frac{\log^2B}{2b} \right).$$
The r.h.s in the first line   of  (\ref{eq:intromultanomaly}) comes from a
regularised trace $ {\rm tr}^B(L(A,B))$ described in (\ref{eq:introtraceQL})
with weight $Q=B$.
The r.h.s in the
second line of  (\ref{eq:intromultanomaly}), which corresponds to $ \log 
  {\cal M}_\zeta(A,B)- {\rm tr}^Q\left(L(A,B)\right)$,   arises from a combination of two
types of local terms; (i)  local residues ${\rm res}_x(\log^2AB),
{\rm res}_x( \log^2A)$ and ${\rm res}_x( \log^2B)$ arising  in
formula (\ref{eq:detzeta})  for the zeta determinants established in
\cite{PS}; (ii)   the local residue ${\rm res}_x(L(A,B)\, \log B)$
arising  in a ``defect formula'' for regularised traces (\ref{eq:PSoperator}) also
established in \cite{PS}, 
applied here to the regularised trace  ${\rm tr}^B(L(A,B))$ with regulator  $B$.
Since the operator $\frac{L(A, B)\,  \log B  }{b} - \frac{\log^2 A\,
  B}{2(a+b)}-\frac{\log^2A}{2a}- \frac{\log^2B}{2b} $ turns out to be 
classical (see Lemma \ref{lem:logsquare}), combining these local
residues yields a 
well-defined noncommutative  residue. \\ \indent
Our approach to the multiplicative anomaly of zeta determinants  is inspired
by  Okikiolu's in \cite{O2}. Before   actually computing the multiplicative anomaly, she first showed \cite{O1}  that for operators $A$ and $B$ with scalar leading symbols,
$$
L(A,B)\simeq \sum_{k=2}^{\infty } C^{(k)}(\log A, \log
  B),
$$
i.e. that  $L(A,B)- \sum_{k=2}^{n +1} C^{(k)}(\log A, \log
  B)$ is of order $<-n$, thus generalising the usual
Campbell-Hausdorff formula to classical pseudodifferential operators with scalar leading
symbols.  Here $C^{(k)}(
\log A, \log B)$ are  Lie monomials given by iterated brackets\footnote{Their
  precise definition is:
 $$C^{(k)}(P,Q):=\frac{1}{k}\, \sum_{j=1}^\infty \frac{(-1)^{j+1}}{j}
 \sum_{\sum_{i=1}^j\alpha_i+\beta_i=k, \, \alpha_j, \beta_j\geq 0}
\frac{({\rm Ad }_P)^{\alpha_1}({\rm Ad }_Q)^{\beta_1}\cdots ({\rm Ad
  }_P)^{\alpha_j}({\rm Ad }_Q)^{\beta_j-1}\,Q}{\,\alpha_1!\cdots\alpha_j!\,
  \beta_1!\cdots\beta_j!}, $$
with the following notational convention: $ ({\rm Ad
  }_P)^{\alpha_j}({\rm Ad }_Q)^{\beta_j-1}\,Q=  ({\rm Ad
  }_P)^{\alpha_j-1}\, P$ if $\beta_j=0$  in which case this vanishes if
  $\alpha_j>1$.}.
\\ \indent Under the assumption that  the operators have scalar leading symbols, the  iterated brackets arising in the Campbell-Hausdorff formula have
 decreasing order, allowing to implement
ordinary traces after a certain order.
In our more general situation, the leading symbols are not necessarily scalar
and  the iterated brackets arising in the
Campbell-Hausdorff formula  hence do not a priori have decreasing order  which
is why we  use
regularised traces instead of the ordinary trace and study
regularised traces of $L(A,B)$.
Okikiolu's proof in the case of operators with scalar leading symbols is
largely based on the observation that the trace of the  operator $L(A,B)- \sum_{k=2}^{n+1 } C^{(k)}(\log A, \log
  B)$ only depends on the
first $n$ positively homogeneous components of $A$ and $B$  where $n$ is the dimension
of the underlying manifold $M$;  this allows her
to work with a finite dimensional space of formal symbols. Interestingly,   regularised  traces of $L(A,B)$  still only depend on the
first $n$ positively homogeneous components of $A$ and $B$. Precisely, given a weight $Q $  and two admissible operators $A$ and $B$ in $\Cl(M, E)$
with non negative orders, we show that   (see Theorem
\ref{thm:trQLABlocal})
\begin{equation}\label{eq:introderivative}\frac{ d}{dt} {\rm tr}^Q(L(A(1+tS), B)= \frac{ d}{dt} {\rm tr}^Q(L(A, B(1+tS))=
0,
\end{equation}
for any operator $S$ in $\Cl(M, E)$  of order $<-n$.
\\   The proofs of Theorems \ref{thm:trQLABlocal}  and \ref{thm:weightedTraceLAB}
 both use the fact that
differentiation in $t$ commutes with  regularised
traces on differentiable families of constant order, a fact we prove  in Proposition
\ref{prop:difftrQ}.  
\\ \indent To conclude, this
approach  sheds light on the locality of
multiplicative anomalies for regularised  determinants  (weighted determinants on the one hand  and
$\zeta$-determinants on the other) in so far as it relates it to the
cyclicity of the noncommutative residue and hence to the multiplicativity of the residue determinant  via
the locality of regularised traces of $L(A,B)$, which are interesting in their
own right. 
\section{The noncommutative residue }
\setcounter{equation}{0}\indent We  recall a few basic definitions concerning classical
pseudodifferential operators
 on closed manifolds, set some notations and define the noncommutative residue
 introduced by Wodzicki in \cite{W1}.\\ \indent
Let U be an open subset of $\R^n$. Given $a \in \C$, the space of
symbols $S^a (U)$  consists of functions $\sigma(x, \xi)$ in
$C^{\infty}(U\times\R^n)$ such that for any compact subset $K$ of
$U$ and any two multiindices $\alpha=(\alpha_1,\cdots,\alpha_n)\in
\N^n, \beta=(\beta_1,\cdots,\beta_n) \in \N^n$ there exists a
constant $C_{K \alpha \beta}$ satisfying for all $(x,\xi)\in
K\times\R^n$
\begin{equation}\label{eq:symbolestimate} \vert \partial_x^{\alpha}\partial_{\xi}^{\beta}\sigma(x,\xi) \vert \leq
 C_{K \alpha \beta}(1+ \vert \xi \vert )^{{\rm Re}(a)- \vert \beta\vert},\end{equation}
where ${\rm Re}(a)$ is the real part of $m$
and $\vert \beta\vert = \beta_1+\cdots+\beta_n.$ \\ \indent
If ${\rm Re}(a_1)<{\rm Re}(a_2)$, then $S^{a_1}(U)\subset S^{a_2}(U)$.\\\\ 
The product $\star$ on symbols is defined as follows:
 if $\sigma_1 \in S^{a_1}(U)$ and $\sigma_2
\in S^{a_2}(U)$,
 \begin{equation}\label{eq:starproduct}
\sigma_1\star \sigma_2(x, \xi) \sim \sum_{\alpha \in
\N^n}\frac{(-i)^{\vert
  \alpha\vert}}{\alpha!}\partial_{\xi}^{\alpha}
\sigma_1(x,\xi)\partial_x^{\alpha}\sigma_2(x,\xi)
\end{equation}
i.e. for any integer $N\geq 1$ we have
$$\sigma_1\star\sigma_2(x,\xi)
-\sum_{\vert \alpha\vert<N}\frac{(-i)^{\vert
\alpha\vert}}{\alpha!}\partial_{\xi}^{\alpha}\sigma_1(x,\xi)\partial_x^{\alpha}\sigma_2(x,\xi)
\in S^{a_1+a_2-N}(U).$$
In particular, $\sigma_1\star \sigma_2 \in S^{a1 + a_2}(U).$\\ \indent
 We denote by $S^{-\infty}(U):= \bigcap_{a \in \C}S^a(U)$  the algebra of
smoothing symbols on $U$, by $S(U):= \langle\bigcup_{a \in
\C}S^a(U)\rangle$ the algebra generated
by all symbols on $U$.  \\ \indent
A symbol $ \sigma$ in $ S^a(U)$ is called  classical of order $a$  if
there is a smooth cut-off function  $\chi\in C^{\infty}(\R^n)$ which
vanishes for
 $\vert \xi \vert \leq {1 \over2}$ and such that $\chi(\xi)=1$ for $\vert \xi
 \vert \geq 1$ such that
\begin{equation}\label{eq:sigmaasympt}\sigma(x, \xi) \sim
\sum_{j=0}^{\infty}\chi(\xi)\, \sigma_{a-j}(x, \xi) \end{equation}
i.e. if  for any integer $N\geq 1$, we have
\begin{equation}\label{eq:sigmaN}\sigma_{(N)}(x, \xi):= \sigma-\sum_{j=1}^{N-1}\chi(\xi)\sigma_{a-j}(x,
  \xi) \in S^{a-N}(U),
\end{equation}
where $\sigma_{a-j}(x, \xi)$ is a positively homogeneous
 function on $U\times\R^n$ of degree $a-j$, i.e. $\sigma_{a-j}(x,
 t\xi)=t^{a-j}\sigma_{a-j}(x, \xi)$ for all $t\in\R^+$.
 \\ \indent
Let $CS^a(U)$ denote the subset of classical symbols of order $a$.
The symbol product of two classical symbols is a classical symbol
and we denote by
$$CS(U)=\langle \bigcup_{a\in \C} CS^a(U)\rangle$$ the algebra
generated by all  classical symbols on $U$.\\\\ \indent The noncommutative  residue of a symbol $\sigma \in
CS(U)$ at point $x$ in $U$
is defined by
 \begin{equation}\label{eq:residuesigma}{\rm res}_x(\sigma):= \int_{S_x^* U} \left(\sigma(x,
  \xi)\right)_{-n}\, \dbar_S \xi,\end{equation}
where $S_x^*U \subset T_x^*U$ is the cotangent unit sphere at point $x$ in $U$,
$\dbar_S\xi=\frac{1}{(2\pi)^n}\, ds\xi$ is the normalised volume measure on the
sphere induced by the canonical volume measure on$\R^n$ and where as
before, $(\cdot)_{-n}$
denotes the
positively homogeneous component  degree $-n$ of the symbol. \\\\\indent 
Given a symbol $\sigma$ in $S(U)$, we can associate to it the
continuous operator $Op(\sigma):C^{\infty}_c (U)\to C^{\infty}(U)$
defined  for $  u\in C^{\infty}_c (U)$-- the space of smooth
compactly supported functions on $U$-- by
$$\left(Op(\sigma)u\right)(x)=\int{e^{ix.\xi}\sigma(x, \xi)\widehat
  u(\xi)\dbar\xi}, $$
where   $\dbar \xi:= \frac{1}{(2\pi)^n}\, d\, \xi$ with $d\xi$ the
ordinary Lebesgue measure on $T_x^*M\simeq \R^n$ and  $\widehat
u(\xi)$
 is the Fourier transform of $u$.
 Since
 $$(Op(\sigma)u)(x)=\int{\int{e^{i(x-y).\xi}\sigma(x, \xi) u(y)\dbar\xi dy}},$$
 $Op(\sigma)$ is an operator with Schwartz kernel given by
$k(x,y)=\int{e^{i(x-y).\xi}\sigma(x, \xi)\dbar\xi},$
which  is smooth off the diagonal.\\
 A pseudodifferential operator $A$  on $U$ is an operator which can be written in
the form $A=Op(\sigma)+R$ where $\sigma $ is a symbol in $ S(U)$
with compact support   and where  $R$ is a smoothing operator  i.e.
$R$ has a smooth kernel.  Its symbol $\sigma_A\sim \sigma$ is
defined modulo smoothing symbols. If $\sigma$ is a classical symbol
of order $a$, then $A$ is called a classical pseudodifferential
operator of order $a$. The composition of two classical operators
$A_1$ and $A_2$ with symbols $\sigma_{A_1}$ and $\sigma_{A_2}$ and
orders $a_1$ and $a_2$ respectively, is a classical operator
$A_1A_2$ of order
$a_1+a_2$ with symbol  $\sigma_{A_1A_2}\sim \sigma_{A_1}\star \sigma_{A_2}$.
\vskip 0,3cm  \indent 
More generally, let $M$ be a smooth  closed  manifold  of dimension
$n$ and $\pi:E\to M$  a smooth finite rank vector bundle over $M$ modelled on
a linear space $V$;
an operator $A:C^{\infty} (M,E)\to C^{\infty}(M,E)$ is a  (resp. classical) pseudodifferential operator of order $a$ if given a local
trivialising chart $(V,\phi)$ on $M$,
 for any localisation
$ A_{\nu}=\chi_{\nu}^2 A\chi_{\nu}^1:C^{\infty}_c (V)\to
C^{\infty}_c (V) $
 of $A$ where $ \chi_{\nu}^i \in C^{\infty}_c (V)$, the operator
$ \phi_*(A_{\nu}):= \phi A_{\nu} \phi^{-1}$ from the space
$C^{\infty}_c (\phi(V))$ into $C^{\infty} (\phi(V))$ is a  (resp.
classical)
pseudodifferential operator of order $a$.\\
 Let $\Cl^a(M, E)$ denote
the set of classical pseudodifferential operators of order $a$.\\
If $A_1\in \Cl^{a_1}(M, E), A_2\in \Cl^{a_2}(M, E)$,
 then the product $A_1 A_2$ lies in $\Cl^{a_1 + a_2}(M, E)$
and we denote by
$$\Cl(M,E):= \langle\bigcup_{a\in \C} \Cl^a(M, E)\rangle$$
the algebra generated by all classical pseudodifferential operators
acting on smooth sections of $E$. Let us also introduce the algebra $$\Cl^{-\infty}(M, E):=
\bigcap_{a\in \R} \Cl^a(M, E)$$
of smoothing operators.
\\ \indent Wodzicki proved that any linear form on the algebra $\Cl(M,
E)$  which vanishes on operator brackets is proportional to the noncommutative
residue \cite{W1}. It is built from
the noncommutative residue density at point $x$ in $M$ defined by
$$\omega_{\rm res}(A)(x):=  {\rm res}_x(\sigma_A)\,
 \, dx; \quad {\rm with}\quad  {\rm res}_x(\sigma_A):=
\int_{S_x^*M}{\rm tr}_x\left(\left(\sigma_A(x,
  \xi)\right)_{-n}\right)\, \dbar \xi,$$
where $S_x^*M \subset T_x^*M$ is the cotangent unit sphere at point $x$ in $M$,
$\dbar_S\xi=\frac{1}{(2\pi)^n}\, ds\xi$ is the normalised volume measure on the
sphere induced by the canonical volume measure on$\R^n$ and where as
before, $(\cdot)_{-n}$
denotes the
positively homogeneous component  degree $-n$ of the symbol. \\
It turns out to  be a globally
defined density on the manifold and gives rise to  the noncommutative residue  introduced in \cite{W1} and  \cite{G}
\begin{equation}\label{eq:resA} {\rm res}(A):= \int_M \omega_{\rm res}(A)(x):= \int_{M} \,\left(\int_{S_x^*M}{\rm tr}_x\left(\left(\sigma_A(x,
  \xi)\right)_{-n}\right)\, \dbar \xi\right)\, dx.
\end{equation}
This linear form which only depends on the $-n$ homogeneous part of the symbol of
the operator, vanishes on operators of order $<-n$ and is local in the sense that
it only depends on a finite number of positively homogeneous components of the
symbol of the operator.\\ \indent By results of Wodzicki \cite{W1} and  Guillemin
\cite{G}  (see also \cite{BG} and \cite{L}),  all traces on $\Cl(M, E)$, i.e. all linear forms
which vanish on commutators in $[\Cl(M, E), \Cl(M, E)]$ are proportional to
the noncommutative residue or equivalently\footnote{Note that the biorthogonal
  (for the dual product) of a subspace $F$ of any vector space  coincides
  with $F$ (see e.g. \cite{B} Chapter II  par. 4 n.6  Proposition 10).} (see
also \cite{L})
\begin{equation}\label{eq:zerores}\forall A\in \Cl(M, E)\quad \left( {\rm res}(A)= 0\Longrightarrow A\in [\Cl(M, E),
\Cl(M, E)]\right).\end{equation}
\section{Logarithms of operators: $\log(AB)-\log A-\log B$}
\setcounter{equation}{0}\indent 
We review the construction and properties of logarithms of elliptic operators
and prove (see Proposition \ref{prop:LAB}) that the expression $\log(AB)-\log A-\log B$ is a finite sum of
commutators of zero order classical pseudodifferential operators.\vskip 0,3cm\indent
 An operator $A
\in \Cl(M,E)$ has principal angle $\theta$ if for every $(x, \xi)\in
T^*M-\{0\}$, the leading symbol $\left(\sigma_A(x, \xi)\right)^L$
has no eigenvalue on the ray  $L_\theta=\{re^{i\theta}, r\geq 0\}$;
in that case $A$ is
elliptic.
\begin{defn} We call an  operator $A\in \Cl(M, E)$
 admissible with spectral cut $\theta$
if $A$ has principal angle $\theta$ and the spectrum of $A$ does
 not meet $  {\text{L}}_{\theta}=\{re^{i\theta}, r\geq  0\}$. In
particular such an operator is  invertible and elliptic. Since the
spectrum of $A$ does not meet $L_{\theta}$, $\theta$ is called an
Agmon angle of $A$.
\end{defn}
\begin{rk}{\rm  In applications, an invertible operator $A$ is often obtained from an essentially self-adjoint elliptic  operator $B\in \Cl(M, E)$ by setting $A=
  B+ \pi_B$   using the orthogonal projection $\pi_B$ onto the kernel $ Ker(B)$ of $B$
  corresponding to the  orthogonal splitting $L^2(M, E)= {\rm Ker}(B)\oplus
{\rm R}(B)$ where $R(B)$ is the (closed) range of $B$. Here $L^2(M, E)$ denotes the closure of $\Ci(M, E)$ w.r. to
a Hermitian structure on $E$ combined with a Riemannian structure on $M$.}
\end{rk}
\indent 
Let $A\in \Cl(M,E)$ be admissible with spectral cut $\theta$ and positive
order $a$. For ${\rm Re}(z)<0$, the complex power $A_{\theta}^z$ of $A$ is
defined by the Cauchy integral
$$A_\theta^{z}=\frac{i}{2\pi} \int_{\Gamma_{r,\theta}} \lambda_\theta^{z}
(A-\lambda)^{-1}\, d\lambda$$
 where $\lambda_\theta^z= \vert
\lambda\vert^z e^{iz {(\rm arg}\lambda)}$ with  $\theta\leq {\rm
arg}\lambda<\theta+2\pi$. In particular,
 for $z=0$, we have $ A_{\theta}^0 = I$.
\\
Here
\begin{equation}\label{eq:contourGamma}\Gamma_{r,\theta}=\Gamma_{r,\theta}^1\cup\Gamma_{r,\theta}^2\cup\Gamma_{r,\theta}^3\end{equation}
where
 $$\Gamma_{r,\theta}^1=\{ \rho \,e^{i\theta}, \infty > \rho\geq r\}$$
 $$\Gamma_{r,\theta}^2=\{ \rho \,e^{i(\theta-2\pi)}, \infty > \rho\geq r\} $$
 $$\Gamma_{r,\theta}^3= \{r\, e^{it},\theta-2\pi\leq t <\theta\},$$
is a contour along the ray $L_\theta$ around the non zero spectrum
of $A$. Here $r$ is any small positive real  number such that
$\Gamma_{r,\theta}\cap Sp(A) =\emptyset.$
  \\
The operator $A_{\theta}^z$ is a classical pseudodifferential operator of order
$az$ with  homogeneous components of the symbol of
$A_\theta^{z}$ given by  $$
\sigma_{az-j}(A_\theta^{z})(x,\xi)=\frac{i}{2\pi}
\int_{\Gamma_\theta} \lambda_\theta^{z}\,b_{-a-j}(x,\xi,\lambda)\,
d\lambda$$ where the components $b_{-a-j}$ are the positive
homogeneous components of the resolvent $(A-\lambda I)^{-1}$ in
$(\xi,\lambda^{\frac{1}{a}})$). In particular,  its leading symbol is given by
$\left(\sigma_{A_{\theta}^z}(x,
\xi)\right)^L=\left((\sigma_{A}(x, \xi))^L\right)_{\theta}^z$ and hence $A_{\theta}^z$ is
elliptic.
\\
The definition of complex powers can be extended to the whole
complex plane by setting $A_{\theta}^z:=A^kA_{\theta}^{z-k}$ for
$k\in \N$ and ${\rm Re}(z)<k$; this definition is independent of the
choice of $k$ in $\N$ and preserves the usual properties, i.e.
$A_{\theta}^{z_1}A_{\theta}^{z_2}=A_{\theta}^{z_1+z_2}, $
$A_{\theta}^k=A^k, {\rm for}\quad k\in\Z.$\\
Complex powers of operators depend on the choice of spectral cut. Wodzicki
\cite{W1}  (Ponge in  \cite{Po1}, see   Proposition
4.1,  further quotes an unpublished paper by
Wodzicki \cite{W2}) established the following result. \begin{prop}\label{prop:spectralcutproj}
\cite{W1,W2,Po1}
Let $\theta$ and $\phi$ be two spectral cuts for an admissible operator $A$ in
$\Cl(M, E)$
 such that
$0\leq \theta <\phi<2\pi$. The complex powers for these two spectral cuts are
related by
\begin{equation}\label{eq:complexpowerspectralcut}
A_\theta^z-A_\phi^z= \left(1-e^{2i\pi z} \right)\, \Pi_{\theta, \phi} (A)
A_\theta^z,
\end{equation} where we have set
$\Pi_{\theta, \phi} (A)= A\, \left( \frac{1}{2i\pi} \int_{\Gamma_{\theta,
      \phi}} \lambda^{-1} (A-\lambda)^{-1} \, d\lambda\right)$ where
$\Gamma_{\theta,
  \phi}$ is a contour around the cone
\begin{equation}\label{eq:cone}
\Lambda_{\theta, \phi}:= \{ \rho \, e^{it}, \infty > \rho\geq r,
\quad \theta<t<\phi\}.
\end{equation}
\end{prop}
\begin{rk}{\rm  Formula (\ref{eq:complexpowerspectralcut}) generalises to spectral cuts $\theta$ and $\phi$ such
  that $0\leq \theta < \phi+2k\pi <(2k+1)\pi$ for some non negative integer
  $k$ by
\begin{equation}\label{eq:kcomplexpowerspectralcut}
A_\theta^z-A_\phi^z= e^{2ik\,\pi z} \,I+ \left(1-e^{2i\pi z} \right)\, \Pi_{\theta, \phi} (A)
A_\theta^z.
\end{equation}}
\end{rk}
If  the cone $\Lambda_{\theta, \phi}$ defined by (\ref{eq:cone})
 delimited by the angles
$\theta$ and $\phi$  does not intersect the
spectrum of the leading symbol of $A$,
 it only contains a
finite number of eigenvalues of $A$ and $\Pi_{\theta, \phi} (A)$ is
a finite rank projection and hence a smoothing operator. In general  (see
Propositions 3.1 and 3.2
in
\cite{Po1}), $\Pi_{\theta, \phi}(A)$, which is a
pseudodifferential projection, is a zero order operator with
leading symbol given by $\pi_{\theta, \phi}(\sigma^L(A))$ defined
similarly to $\Pi_{\theta, \phi}$ replacing $A$ by the leading
symbol $\sigma_A^L$ of $A$ so that:
$$\sigma^L_{\Pi_{\theta, \phi} (A)}=\pi_{\theta, \phi} (\sigma^L_A) := \sigma^L_A\,
\left( \frac{1}{2i\pi} \int_{\Gamma_{\theta, \phi}} \lambda^{-1}
(\sigma_A^L-\lambda)^{-1} \, d\lambda\right),$$
where we have set $\sigma_B^L(x, \xi)= \left(\sigma_B(x, \xi)\right)^L \quad
$ for any $ (x, \xi)\in T^*M$ and any $B\in Cl(M, E)$.  \\
We are finally ready to define the logarithm.
The logarithm  of an admissible operator $A$ with spectral cut $\theta$  is defined in terms of the derivative at
$z=0$
of this complex power:
$$\log_\theta(A)=  \partial_z {A_\theta^{z}} _{\vert_{z=0}}.$$
\begin{rk}{\rm  For a real number $t$, $A$ and $A_\theta^t$ have spectral cuts $\theta$ and
$t\theta$; for $t$ close to one,
$(A_\theta^t)_{t\theta}^z=(A_\theta^t)_{\theta}^z$ and hence,
$(A_\theta^t)_{t\theta}^z=A_\theta^{tz}$ so that
\begin{equation*}\label{eq:logAt}
\log_{\theta}(A^t)= \partial_z {(A_\theta^t)_{t\theta}^{z}}
_{\vert_{z=0}}
=\partial_z {(A_\theta^{tz})} _{\vert_{z=0}}= t\log_\theta A.
\hspace{3cm}
\end{equation*}}
\end{rk}
Just as complex powers, the  logarithm depends on the choice of
spectral cut \cite{O1}. Indeed, differentiating
(\ref{eq:complexpowerspectralcut}) w.r. to $z$ at $z=0$  yields for spectral
cuts $\theta, \phi$ such that  $0\leq \theta <\phi <2\pi$  (compare with  formula (1.4) in \cite{O1}):
\begin{equation}\label{eq:logspectralcut}
\log_\theta A-\log_\phi A= -2i\pi \Pi_{\theta, \phi}(A).
\end{equation}
 Formula (\ref{eq:logspectralcut}) generalises to spectral cuts $\theta$ and $\phi$ such
  that $0\leq \theta < \phi+2k\pi <(2k+1)\pi$ for some non negative integer
  $k$ by
\begin{equation}\label{eq:klogspectralcut}
\log_\theta A-\log_\phi A= 2ik\,\pi  \,I-2i\pi \Pi_{\theta, \phi}(A).
\end{equation}
As a result of the above discussion and as already observed in
\cite{O1}, when the leading symbol $\sigma^L_A$ has no eigenvalue
inside the cone $\Lambda_{\theta, \phi}$ delimited by
$\Gamma_{\theta, \phi}$ then $\Pi_{\theta, \phi}$ which is a finite
rank projection, is smoothing.
\vskip 0,3cm \indent Logarithms of
classical pseudodifferential operators are not classical anymore since their
symbols involve a logarithmic term $\log\vert \xi\vert$ as the following
elementary result
shows.
\begin{prop}\label{prop:leadsymbolsigmathetaA}
 Let $A\in \Cl(M, E)$ be an admissible operator with spectral cut $\theta$.
In a local trivialisation, the symbol of $\log_\theta (A)$ reads:
\begin{equation}\label{eq:symbollog}
\sigma_{\log_\theta(A)}(x, \xi)=a\, \log \vert \xi\vert I +
\sigma_0^A(x, \xi)
\end{equation}
 where $a$ denotes the order of $A$ and $\sigma_0^A$ a symbol of
order zero.\\
Moreover, the leading symbol of $\sigma_0^A$ is given by
 \begin{equation}\label{eq:leadinglogsymb}(\sigma_0^A)^L(x, \xi)=\log_\theta
\left(\sigma_{A}^L(x,\frac{\xi}{\vert
  \xi\vert})\right)\quad \forall (x, \xi)\in T^*M-\{0\}.\end{equation}
In particular, if $\sigma_A$ has scalar leading symbol then so  have
 $\sigma_\theta^A$ and $\sigma_{\Pi_{\theta, \phi} (A)} $ for any other
 spectral cut $\phi$.
\end{prop}
{ \bf Proof:}
Given a local trivialisation over some local chart, the symbol of $A_\theta^z$ has
the formal expansion $\sigma_{A_\theta^z} \sim \sum_{j \geq
0}b^{(z)}_{az-j} $ where $a$ is the order of $A$ and
$b^{(z)}_{az-j}$ is a positively homogeneous function of degree $az-j$. Since
$\log_\theta A= \partial_z {A_\theta^{z}}_{ \vert_{z=0}}$, the
formal expansion of the symbol of
 $\log_\theta A$ is $\sigma_{\log_\theta A}\sim \sum_{j \geq 0}\partial_z
{b^{(z)}_{az-j}}_{\vert_{z=0}} $\\
Since   ${A_\theta ^z}_{ \vert_{z=0}}=I$, we have $\sigma_{{A_\theta^{z}}_{ \vert_{z=0}}}
\sim I$ where now  $I$ stands for the identity on matrices. Thus  $b^{(z)}_{az-j}(x,
\xi)_{ \vert_{z=0}}= \delta_{0,j}I.$\\
Suppose that $\xi \ne 0$; using the positive homogeneity of the components, we
have:
$b^{(z)}_{az}(x,\xi) = \vert \xi\vert^{az} b^{(z)}_{az}\left(x,\frac{\xi}{\vert
  \xi\vert}\right);$
hence $$ \partial_z b^{(z)}_{az}(x,\xi)=
a \log\vert \xi\vert {\vert \xi\vert}^{az} b^{(z)}_{az}\left(x,\frac{\xi}{\vert
  \xi\vert}\right)+
{\vert \xi\vert}^{az} \partial_z b^{(z)}_{az}\left(x,\frac{\xi}{\vert
  \xi\vert}\right).$$
It follows that
$$ \partial_z b^{(z)}_{az}(x,\xi)_{\vert_{z=0}}=a \log\vert \xi\vert I+
\partial_z b^{(z)}_{az}\left(x,\frac{\xi}{\vert
  \xi\vert}\right)_{\vert_{z=0}}.$$
Similarly for $j>0$, we have
$\quad b^{(z)}_{az-j}(x,\xi)= \vert \xi\vert^{az-j}
b^{(z)}_{az-j}\left(x,\frac{\xi}{\vert\xi\vert}\right)$ so that
$$
\partial_z b^{(z)}_{az-j}(x,\xi)= a\log\vert \xi\vert {\vert
\xi\vert}^{az-j}b^{(z)}_{az-j}\left(x,\frac{\xi}{\vert
  \xi\vert}\right)+ \vert \xi\vert^{az-j} \partial_z
b^{(z)}_{az-j}\left(x,\frac{\xi}{\vert
  \xi\vert}\right).$$
 Consequently,
$$\partial_z b^{(z)}_{az-j}(x,\xi)_{\vert_{z=0}}=
\vert \xi\vert^{-j} \partial_z b^{(z)}_{az-j}\left(x,\frac{\xi}{\vert
  \xi\vert}\right)_{\vert_{z=0}}.$$
  The terms $\partial_z
b^{(z)}_{az}(x,\frac{\xi}{\vert
  \xi\vert})$ and $\partial_z b^{(z)}_{az-j}(x,\frac{\xi}{\vert
  \xi\vert})$ are homogeneous functions of  degree 0 in $\xi$.
 Summing up, we obtain
$$\sigma_{\log_\theta(A)}(x, \xi)=a\, \log \vert \xi\vert I + \sigma_\theta^A(x,
\xi) $$
where
$\sigma_0^A(x, \xi)= \partial_z b^{(z)}_{az}(x,\frac{\xi}{\vert
  \xi\vert})_{\vert_{z=0}}+\sum_{j>0}\vert \xi\vert^{-j} \partial_z
b^{(z)}_{az-j}(x,\frac{\xi}{\vert
  \xi\vert})_{\vert_{z=0}}.$
$\sigma_0^A$ is a symbol of order 0.  Its leading symbol reads
$(\sigma_0^A(x, \xi))^L=\partial_z b^{(z)}_{az}(x,\frac{\xi}{\vert
  \xi\vert})_{\vert_{z=0}}=\partial_z
{\left(\sigma_{A}^L(x,\frac{\xi}{\vert
  \xi\vert}\right)_\theta^{z}}_{ \vert_{z=0}}=\log_\theta
\sigma_{A}^L(x, \frac{\xi}{\vert
  \xi\vert})$ for any $ (x, \xi)$ in $T^*M-\{0\}$. \endsquare\\\\ \indent This
motivates the introduction of log-polyhomogeneous symbols (see
e.g. \cite{L}), to which the local noncommutative residue easily extends.
\begin{defn} A symbol $\sigma \in S(U)$ is called log-polyhomogeneous of order
  $a$ and type
  $k$ for some non negative integer $k$  if there is some smooth function
  $\chi$ on $\R^n$  which vanishes around zero and is identically one outside the unit
  ball, such that
$$\sigma(x, \xi) \sim
  \sum_{j=0}^{\infty}\chi(\xi)\, \left(\sigma(x,  \xi)\right)_{a-j}$$
  where for any non negative integer $j$,
 $$\left(\sigma_A(x, \xi)\right)_{
  a-j}= \sum_{l=0}^k \left(\sigma_A(x, \xi)\right)_{a-j,l}(x,
  \xi)\, \log^l \vert \xi\vert \quad\forall (x, \xi)\in T^*U,$$
with  $\sigma_{
  a-j,l}, l=0,\cdots, k$ are positively homogeneous of degree $a-j$.
\\ The local noncommutative residue at a point $x$ in $U$   defined in (\ref{eq:residuesigma}) extends
to log-polyhomogeneous symbols by:
\begin{equation}\label{eq:residuelogsigma}{\rm res}_x(\sigma):= \int_{S_x^* U} \left(\sigma(x,
  \xi)\right)_{-n}\, \dbar_S \xi.\end{equation}
\end{defn}\indent 
Powers of
the logarithm of a given admissible operator combined with all
classical pseudodifferential operators generate the algebra of
log-polyhomogenous operators  \cite{L}. A
log-polyhomogenous operator $A$ of type $k$ is a pseudodifferential operator whose local
symbol $\sigma_A(x, \xi)$ in any local trivialisation asymptotically is
log-polyhomogeneous of type $k$.\\
 Let us  denote the set of
such operators by $\Cl^{a, k}(M,E)$ and  its union over all non
negative integers $k$ by $\Cl^{a, *}(M,E)=\cup_{l=0}^k \Cl^{a, l}(M,
E)$.  In particular, a classical pseudodifferential operator is a
log-polyhomogeneous operator of log-type $0$ and $\Cl^{a, 0}(M, E)=
\Cl^a(M, E)$. The product of a log-polyhomogeneous operator of type
$k$ and a log-polyhomogeneous operator of type $l$ is
log-polyhomogeneous operator of type $k+l$ so that, following
\cite{L}, we can build  the algebra $\Cl^{*, *}(M, E)=\langle
\cup_{a\in\C, k\in \Z_+}\Cl^{a, k}(M, E)\rangle$  generated by all
log-polyhomogeneous operators. \\ For  an operator $A$ in $ \Cl^{*, *}(M, E)$,
one can define the local  noncommutative residue at a point $x$ in $M$
similarly to the case of classical operators
 by:  $${\rm res}_{x}(A):=
\int_{S^*_xM}{\rm tr}_x\left( \sigma_A(x, \xi)\right)_{-n}\, \dbar_S
\xi.$$ However, unlike Lesch's extended noncommutative residue on
log-polyhomogeneous
  operators \cite{L}, the locally defined residue density  ${\rm res}_x(A)\, dx$
  is not expected to  patch  up to a globally defined
  residue density.
\\ However, it does for logarithms of any admissible operator $A$ in $\Cl(M,
E)$  and we have \cite{Sc}:
 \begin{equation}\label{eq:reslog}{\rm res}(\log A)= -a \, \zeta_A(0)\end{equation}
where $\zeta_A(0)$ is the constant term in the Laurent expansion of the unique
meromorphic extension $\zeta_A(s)$  of the map $s\mapsto {\rm
  Tr}(A^{-s})$ given by the $L^2$-trace of $A^{-s}$ defined on the domain of
holomorphicity  ${\rm Re}(as)>n$ \footnote{This actually is an instance in the
  case $A(z)= A^{-z}$ of the
  more general defect formula (\ref{eq:PSoperator}) derived in  \cite{PS}.}.\\
In \cite{Sc}, Scott showed the multiplicativity of the associated residue determinant $${\rm
  det}_{\rm res}(A):= e^{{\rm
  res}(\log A)}.$$  He actually proved a stronger statement, namely that  given  two admissible  operators $A,B$ such that their
  product $AB$ is also admissible, the  following expression
$$L(A, B):=\log (AB)- \log A-\log B$$
has vanishing noncommutative residue.  \begin{rk}\label{rk:spectralcutL}{\rm Strictly speaking, one should specify the spectral cuts $\theta$ of
  $A$, $\phi$ of $B$ and $\psi$ of $AB$ in the expression $L(A,B)$  setting instead
$$L^{\theta, \phi, \psi}(A, B):=\log_\psi (AB)-\log_\theta
  A-\log_\phi B.$$
Then by (\ref{eq:logspectralcut})
$$L^{\theta, \phi, \psi}(A, B)-L^{\theta^\prime, \phi^\prime, \psi^\prime}(A,
B)=-2i\pi \, \left(\Pi_{\psi, \psi^\prime}(AB)- \Pi_{\theta,
    \theta^\prime}(A)-\Pi_{\phi, \phi^\prime}(B)\right).$$ Since these have
vanishing residue  by results of Wodzicki \cite{W1}, a change of spectral cut
does not affect the residue of $L(A,B)$. \\
Up to a modification of the operators $A$ and $B$, one  can actually  choose fixed spectral cuts $\theta$
and $\phi$ by the following argument of Okikiolu \cite{O1}:
$$ L^{\theta, \phi, \psi}(A, B)= L^{\pi, \pi,
\psi-(\theta+\phi)}(e^{i(\pi-\theta)}A, e^{i(\pi-\phi)}B).$$
Indeed, if $A, B, AB$ have spectral cut $\theta, \phi, \psi$ respectively, then
 $A^\prime=e^{i(\pi-\theta)}A$ and $B^\prime=e^{i(\pi-\phi)}B$ have spectral cut
 $\pi$ and $A^\prime B^\prime$ has spectral cut $\psi+2\pi-\theta-\phi$.
So we can assume that $\theta=\phi=\pi$ without loss of generality.\\
 Keeping in mind these observations, in order to
simplify notations we assume that $A$ and $B$ have spectral cuts $\pi$ and  drop the explicit mention of the spectral cuts.}\end{rk}
It follows from (\ref{eq:zerores}) that
  \begin{equation}\label{eq:Lbracket} L(A,B)\in [\Cl(M, E), \Cl(M, E)], \end{equation} so that  $L(A,B)$  is a finite sum
of commutators.  The following proposition  provides a refinement this statement.
\begin{prop} \label{prop:LAB}
Let $A, B$ be two admissible  operators, which w.l.o.g. are
assumed   to have $\pi$ as
spectral cut (see Remark \ref{rk:spectralcutL}), such that their
  product $AB$ is also admissible with spectral cut $\pi$.  Then
$L(A, B)$ is a finite sum of Lie brackets of operators in $\Cl^0(M, E)$:
$$ L(A, B) \in [\Cl^0(M, E), \Cl^0(M,
E)].$$
\end{prop}
{\bf Proof:}  Let us  check that $L(A,B)$ lies in $\Cl^0(M, E)$.  If $Â$ has
order $a$ and $B$ has order $b$ then  $AB$ has order $a\,+b$, we have
 \begin{eqnarray}\label{eq:sigmaLAB}
   \sigma_{L(A,B)}&=& \sigma_{\log AB}(x, \xi)-\sigma_{\log A}(x, \xi)-\sigma_{\log B}(x, \xi)\nonumber\\
    &=&(a+b)\,\log\vert \xi\vert\,  I+\sigma_0^{AB} (x,\xi)-a\,\log\vert \xi\vert I\nonumber\\
    &{}&- \sigma_0^A (x,\xi) -b\,\log\vert \xi\vert\,  I-\sigma_0^B
    (x,\xi)\nonumber\\
    &\sim& \sigma_0^{AB} (x,\xi)-\sigma_0^A (x,\xi)-\sigma_0^B (x,\xi)
  \end{eqnarray}
 so that the operator $L(A,B)$ is indeed classical of order $0$ and by
 (\ref{eq:leadinglogsymb}) it has  leading
 symbol given for any $(x,\xi)\in T^*M/M$ by
$$\left(\sigma_{L(A, B)}(x, \xi)\right)_0=\log
\sigma_{AB}^L(x, \frac{\xi}{\vert
  \xi\vert})-\log
\sigma_{A}^L(x, \frac{\xi}{\vert
  \xi\vert})-\log
\sigma_{B}^L(x, \frac{\xi}{\vert
  \xi\vert})=: L(\sigma_A^L, \sigma_B^L)\left(x, \frac{\xi}{\vert \xi\vert}\right).$$
Here as before, $\sigma_C^L$ stands for the leading symbol of the operator
$C$. \\
Applying  the usual Campbell-Hausdorff formula to the matrices  $\sigma_A^L\left(x,
  \frac{\xi}{\vert \xi\vert}\right)$ and  $\sigma_B^L\left(x,
  \frac{\xi}{\vert \xi\vert}\right)$ and implementing the fibrewise  trace ${\rm tr}_x$  yields:
$$ {\rm tr}_x\left(
\log\sigma_{AB}^L(x, \frac{\xi}{\vert
  \xi\vert})-\log
\sigma_{A}^L(x, \frac{\xi}{\vert
  \xi\vert})-\log
\sigma_{B}^L(x, \frac{\xi}{\vert
  \xi\vert})\right)={\rm tr}_x\left( L(\sigma_A^L, \sigma_B^L)\left(x, \frac{\xi}{\vert \xi\vert}\right)\right)=0.$$
It follows that any leading symbol trace   ${\rm
  Tr}_0^\Lambda(C):= \Lambda\left(\left({\rm
        tr}_x(\sigma_C)\right)_0\right)$
(see e.g.  \cite{LP}) on
the algebra $\Cl^0(M, E)$ where $\Lambda$ is a current in $ C^\infty(S^*M)'$
and the index $0$ stands for the positively homogeneous component of degree $0$, vanishes on $L(A,B)$:
$${\rm Tr}^\Lambda(L(A,B))= \Lambda\left({\rm tr}_x\left( \sigma_{L(A,
    B)}\right)_0\right)=0.$$
Thus both the noncommutative residue and leading symbol traces vanish on
$L(A,B)$.  But by the results of \cite{LP},
any  trace on $\Cl^0(M, E)$, i.e. any linear form on  $\Cl^0(M, E)$ which vanishes
 on $[ \Cl^0(M, E),  \Cl^0(M, E)]$, is a linear combination of the
 noncommutative residue and a leading symbol trace. Consequently (see
 e.g. \cite{Po2} Corollary 4.5),  all traces
 on $\Cl^0(M, E)$ vanish on the
 operator $L(A,B)$ which therefore
 lies in $[ \Cl^0(M, E),  \Cl^0(M, E)]$.\endsquare

\section{Properties  of  weighted traces }
\setcounter{equation}{0} \indent Since  traces on $\Cl(M, E)$ are
proportional to the noncommutative residue which vanishes on
smoothing operators, the $L^2$-trace on smoothing operators does not
extend to the whole algebra $\Cl(M, E)$. Instead
 we use  linear extensions which we call weighted traces, of the ordinary
 $L^2$-trace on smoothing operators to the whole algebra $\Cl(M, E)$. We
 review basic properties of weighted traces and prove
 (see Proposition \ref{prop:difftrQ}) that
  the canonical and weighted traces as well as the noncommutative residue commute with
 differentiation on differentiable families of operators with constant order.\\\\
 Weighted traces arise as finite parts of canonical traces of holomorphic
 families of classical pseudodifferential operators.\\ A
 family $\{f(z)\}_{z\in \Omega}$  in a topological vector space ${\cal A}$ which is
 parametrised by a complex domain $\Omega$, is holomorphic at
$z_0\in \Omega$ if
the corresponding function $f:\Omega\to{\cal A}$ admits a  Taylor expansion in a neighborhood $N_{z_0}$ of $z_0$
\begin{equation}\label{e:Texpansion}
f(z) = \sum_{k=0}^{\infty}f^{(k)}(z_0)\,\frac{(z-z_0)^k}{k!}
\end{equation}
which is convergent, uniformly on compact subsets in a neighborhood of $z_0$
(i.e. locally uniformly),
with respect to the  topology on ${\cal A}$. The vector space of functions we
consider here is $C^\infty( U\times \R^n)\otimes {\rm End}(V)$  equipped with
the uniform convergence of all derivatives on
compact subsets.
 \begin{defn}\label{defn:holosymb}
 Let $\Omega$ be a domain of $\C$ and $U$ an open subset of $\R^n$.
A  family $(\sigma(z))_{z\in \Omega}$  is a holomorphic family of
End$(V)$-valued  classical
symbols on $U$ parametrised by $\Omega$ when
\begin{enumerate}
\item the map $z\mapsto \alpha(z)$  with  $\alpha(z)$ the order of $\sigma(z)$, is holomorphic in $z$, 
\item  $z\mapsto\sigma(z)$  is holomorphic as element of
  $C^\infty( U\times \R^n)\otimes {\rm End}(V)$ and for each $z\in \Omega$,
  $\sigma(z)\sim \sum_{j=0}^\infty \chi\, \sigma(z)_{\alpha(z)-j}$ (for some
  smooth function $\chi$ which is identically one outside the unit ball and vanishes in a
  neighborhood of $0$) lies in $CS^{\alpha(z)}(U)\otimes {\rm End}(V)$,
\item  for any positive integer $N$, the remainder term
$\sigma_{(N)}(z)= \sigma(z)- \sum_{j=0}^{N-1} \sigma(z)_{\alpha(z)-j}$ is
holomorphic in $z\in \Omega$ as an element of  $C^\infty( U\times
\R^n)\otimes {\rm End}(V)$ and its  $k$-th derivative
 $$(x, \xi)\mapsto  \partial_z^k\sigma_{(N)}(z)(x,\xi):=
 \partial_z^k\left(\sigma_{(N)}(z)(x,\xi)\right)$$
 lies in $ S^{\alpha(z)-N+\e}(U)\otimes
 {\rm End}(V)$ for all
 $\e>0$  locally uniformly in $z$, i.e the $k$-th derivative
$ \partial_z^k\sigma_{(N)}(z) $
 satisfies a uniform estimate (\ref{eq:symbolestimate}) in $z$   on compact
 subsets  in $\Omega$.
\end{enumerate}
In particular, for any integer $j\geq 0,$ the (positively) homogeneous component
$\sigma_{\alpha(z)-j}(z)$ of degree $\alpha(z)-j$ of the symbol  is
holomorphic on $\Omega$ as an element of  $C^\infty( U\times
\R^n)\otimes {\rm End}(V)$.
\end{defn}\indent 
It is important to observe that the derivative  of  a holomorphic family $\sigma(z)$ of classical
  symbols is not classical anymore since it  yields a holomorphic family of symbols $\sigma^\prime(z)$  of
order $\alpha(z)$, the  asymptotic expansion of
  which involves a logarithmic term and reads \cite{PS}:
\begin{equation*}\label{eq:sigmaprime}
\sigma^\prime(z)(x, \xi)\sim\sum_{j=0}^\infty  \chi(\xi)\left( \log \vert
  \xi\vert\, \sigma_{\alpha(z)-j, 1}^\prime(z)(x, \xi)
+ \sigma^\prime_{\alpha(z)-j, 0}(z)(x, \xi)\right)\quad\forall (x, \xi)\in
T^*U
\end{equation*} for some smooth cut-off function $\chi$ around the origin
which is
identically equal to $1$ outside the open unit ball and positively
homogeneous symbols
\begin{equation*}\label{eq:sigmaprimej0}
\sigma_{\alpha(z)-j, 0}^\prime(z)(x, \xi)=\vert \xi\vert^{\alpha(z)-j}\,  \partial_z
\left(\sigma_{\alpha(z)-j}(z)(x,\frac{\xi}{\vert
  \xi \vert})\right),
\end{equation*}
\begin{equation*}\label{eq:sigmaprimej1}
\sigma^\prime_{{\alpha(z)-j}, 1}(z)(x, \xi)=\alpha^\prime(z)\,
\sigma_{\alpha(z)-j}(z)(x,\xi)
\end{equation*}
of degree $\alpha(z)-j$.\\ \indent The regularised cut-off integral on symbols we are
about to introduce is an
essential ingredient to build linear extensions of the $L^2$-trace. \\
The integral $\int_{B_x(0, R)} {\rm
  tr}(\sigma(x, \xi))\dbar \xi $ of the  trace of a symbol
$\sigma\in CS(U)\otimes {\rm End}(V) $  over the ball $B_x(0,R )$
of radius $R$ centered at $0$  in the cotangent space $ T_x^*U$ at a point $x\in U$, has an  asymptotic expansion in
decreasing powers of $R$ which is polynomial in  log $R$ so that the cut-off
integral which corresponds to the constant term in this expansion $$\cutoffint_{T_x^*U}{\rm
  tr}\left(\sigma(x, \xi) \right)\,\dbar \xi := {\rm fp}_{R\to
\infty}\int_{B_x(0, R)} {\rm
  tr}\left(\sigma(x, \xi)\right)\,\dbar \xi$$ is well defined. It coincides with
the
ordinary integral whenever the latter converges.\\
We now recall the properties of  cut-off integrals of holomorphic families of symbols.
 \begin{prop}\label{prop:KVPSsymbol}\begin{enumerate}\item \cite{KV}  The cut-off regularised
 integral $\cutoffint_{T_x^*U} {\rm tr}_x\left( \sigma(z) (x, \xi)\right)\,
 \dbar\xi$  of a  holomorphic family  $\sigma(z)$ of classical pseudodifferential symbols  on a neighborhood
 $U\subset M$ of holomorphic order $\alpha(z)$  is a  meromorphic function in
 $z$ with simple poles. The
 residue at a pole $z_0$ for which $\alpha^\prime(z_0)\neq 0$ is given by:   \begin{equation}\label{eq:KVsymbol}{\rm Res}_{z=z_0} \cutoffint_{T_x^*U} {\rm tr}_x\left( \sigma(z) (x, \xi)\right)\,
 \dbar\xi=-\frac{1}{\alpha^\prime (z_0)} {\rm res}(\sigma(z_0)).
\end{equation} \item \cite{PS} Furthermore, if its holomorphic order is affine and  non constant \footnote{Here and in
   what follows, we assume that
   the order  of the holomorphic family is affine and  non constant  so that applying the
   fibrewise trace, formula (1.50) of
   \cite{PS} boils down to the following one.},  its
 Laurent expansion has  constant
 term at $z_0$ given by
\begin{equation}\label{eq:PSsymbol} {\rm fp}_{z=z_0} \cutoffint_{T_x^*U}
    {\rm tr}_x\left( \sigma(z) (x, \xi)\right)\, \dbar\xi= \cutoffint_{T_x^*U}
    {\rm tr}_x\left(\sigma(z_0)(x, \xi)\right)\,
     \dbar\xi-\frac{1}{\alpha^\prime(z_0)}  {\rm
       res}_x\left(\sigma^\prime
(z_0)\right).\end{equation} Here we use the local residue extended
to log-polyhomogeneous symbols (see (\ref{eq:residuelogsigma})) since the derivative
$\sigma^\prime(z_0)$ of a holomorphic family of classical symbols
$\sigma(z)$ with order $\alpha(z)$ at a point $z_0$ is expected to
be logarithmic with same order.
\end{enumerate}
\end{prop}
Let us  now carry out these constructions to the operator level. \\
 For any $A\in \Cl(M, E)$, for
any $x\in M$, the following expression defines a local density:
\begin{equation}\label{eq:omegaKV}\omega_{KV}(A)(x):= \left(\cutoffint_{T_x^*M} {\rm tr}_x\left(\sigma_A(x,
  \xi)\right)\, \dbar \xi\right)\, dx.
\end{equation}
It patches up to a global density on $M$  whenever the  operator $A$ in $\Cl(M,
E)$ has non integer order or has order $<-n$ so that the Kontsevich-Vishik canonical trace
\cite{KV} (see also \cite{L}):
\begin{equation}\label{eq:TRKV}{\rm TR}(A):= \int_M \omega_{KV}(A)(x):= \int_M
  {\rm TR}_x(A) \, dx
\end{equation}
makes sense.\\
The canonical trace can  be applied to  holomorphic families of classical
pseudodifferential operators with varying complex order.
\begin{defn} \label{defn:holomap} Let  $(A(z))_{z \in \Omega}$ be a family  of
  classical pseudodifferential operators in $\Cl(M, E)$ with distribution
  kernels $(x,y) \mapsto K_{A(z)}(x, y)$.
The family is holomorphic if  
\begin{enumerate}
\item the order $\alpha(z)$ of $A(z)$ is holomorphic in $z$, 
\item      in any local trivialisation of $E$, we can write $A(z)$ in the form
 $A(z)=Op(\sigma_{z})+R(z)$, for some holomorphic family of End$(V)$-valued
 symbols
 $(\sigma(z))_{z\in \Omega}$ where $V$ is the model space of the fibres of  $E$, and some holomorphic family $(R(z))_{z \in
   \Omega}$ of smoothing operators i.e. given by a holomorphic family of
 smooth Schwartz kernels,
\item  the (smooth) restrictions of the distribution kernels $K_{A(z)}$ 
to the  complement of the diagonal $\Delta\subset M\times M$, form a  holomorphic
family 
with respect to the  topology given by  the uniform convergence in all
derivatives on compact subsets of $ M\times M-\Delta$.
\end{enumerate}
\end{defn}
\begin{ex} Given an admissible operator $A\in \Cl(M, E)$ with spectral cut $\theta$, a  family $z \mapsto A^{-\frac{z}{1+\mu z}}_\theta$ for $\mu
\in \R$ is a holomorphic family of $ \psi DOs .$ In particular,
$A(z)=A_\theta^{-z}$ is a holomorphic family. Whereas $A^\prime(z)=\partial_z A_\theta^{-z}$ is classical,
$A^\prime(0)=(\partial_z A_\theta^{-z})_{z=0}=-\log_{\theta}A $ is not.
\end{ex}
Integrating the formulae in Proposition \ref{prop:KVPSsymbol} along the manifold $M$
yields the following result on the level of operators.
\begin{prop}\label{prop:KVPSoperator}\begin{enumerate}\item \cite{KV}  The
    canonical trace ${\rm TR}(A(z))$  of a  holomorphic family  $A(z)$ of
    classical pseudodifferential operators  in $\Cl(M, E)$ of holomorphic
    order $\alpha(z)$  is  a meromorphic function in $z$ with simple poles  and
 residue at a pole $z_0$ for which $\alpha^\prime(z_0)\neq 0$ is given by:   \begin{equation}\label{eq:KVoperator}{\rm Res}_{z=z_0} {\rm TR}(A(z))=-\frac{1}{\alpha^\prime (z_0)} {\rm res}(A(z_0)).
\end{equation} \item \cite{PS} Furthermore, if its holomorphic order is affine and  non constant,  its
 Laurent expansion has  constant
 term at $z_0$ given by
\begin{equation}\label{eq:PSoperator}
{\rm fp}_{z=z_0}{\rm TR}(A(z))= \int_M dx\left( {\rm
    TR}_x(A(z_0))-\frac{1}{\alpha^\prime(z_0)}{\rm res}_x(A^\prime(z_0))\right).
\end{equation}
\end{enumerate}
\end{prop}   Applying these results to a holomorphic family $A(z):=
A\, Q^{-z}$ where $A$ is any operator in $\Cl(M, E)$ and  $Q$  an admissible
operator in $\Cl(M, E)$ with positive order and  spectral cut $\alpha$, we infer  (see e.g. \cite{KV}, \cite{P}, \cite{CDMP} and references therein) that  the
map $z\mapsto {\rm TR}\left(A\, Q_\alpha^{-z}\right)$ is
meromorphic with simple poles.
\begin{defn}\label{defn:weightedtrace} Given an admissible operator $Q$ with positive order, which we call
  a {\rm weight}, the  $Q$-weighted trace of an operator $A$ in $\Cl(M, E)$ is given
by:
$${\rm tr}_\alpha^Q(A):= {\rm fp}_{z=0} {\rm TR}\left(A\,
Q_\alpha^{-z}\right):=  \lim_{z\to 0} \left({\rm TR}\left(A\,
Q_\alpha^{-z}\right)-{\rm Res}_{z=0}\left(\frac{{\rm TR}\left(A\,
Q_\alpha^{-z}\right)}{z}\right)\right),$$ where  $\alpha$ is a
spectral cut for $Q$ and where  ${\rm fp}_{z=0}$ denotes the
constant term in the Laurent expansion.
\end{defn}
Applying  (\ref{eq:PSoperator})  to the family $A(z)= A\, Q_\alpha^{-z}$
yields the following ``defect formula'' \cite{PS}:
\begin{equation}\label{eq:trQPS}{\rm tr}_\alpha^Q(A)=  \int_{M} \left({\rm TR}_x\left(A\right)-\frac{{\rm res}_x\left(A\,
\log_\alpha Q\right)}{q}\right)\, dx, \end{equation}
where $q$ stands for the order of $Q$. In particular, for $A=I$ we get
back (\ref{eq:reslog}):
$$\zeta_{Q,\alpha}(0):= {\rm fp}_{z=0} {\rm TR}(Q_\alpha^{-z})= -\frac{{\rm
    res}\log_\alpha Q}{q}.$$\\
\indent Whereas
weighted traces are not expected to be local in general since they involve the
whole symbol of the operator,   the difference of two weighted traces is local in so far as it
involves a finite number of homogeneous components of the symbol via the
noncommutative residue.
Weighted traces depend on the choice of weight and are not cyclic in
spite of their name.
\begin{prop} \label{prop:propertiestrQ} (\cite{CDMP}, \cite{MN})
\begin{enumerate}
\item  Given two weights  $Q_1$ and $Q_2$ with common spectral
cut $\alpha$ and positive orders $q_1$, $q_2$ we have
\begin{equation} \label{eq:differenceWeighTrace}
{\rm tr}_\alpha^{Q_1}(A)-{\rm tr}_\alpha^{Q_2}(A)= {\rm res}\left(A\,
  \left(\frac{\log_\alpha Q_2}{q_2}  -\frac{\log_\alpha Q_1}{q_1}
  \right)\right),
\end{equation}
which is a  local expression.
\item For any weight $Q$ in $\Cl(M, E)$
with order $q$ and spectral cut $\alpha,$  the operators $ [A,
\log_\alpha Q] $ and $ [B,
\log_\alpha Q] $ lie in $ \Cl(M, E)$ and
\begin{equation}\label{eq:coboundary}
{\rm tr}_\alpha^Q\left( [A, B]\right)=\frac{1}{q} {\rm res} \left( A \, [B,
\log_\alpha Q]\right)=-\frac{1}{q} {\rm res} \left( B \, [A, \log_\alpha Q]\right).
\end{equation}
In particular, if $Q=A$ or $Q=B$, or if the sum of the orders of   $A$ and $B$
has real part $<-n$,  then
 ${\rm tr}_\alpha^Q\left( [A, B]\right)=0.$
\end{enumerate}
\end{prop}
The following technical proposition shows that the canonical and  weighted traces  as
well as the noncommutative residue  commute with
differentiation on families of  operators of constant order, a fact that we will use to
derive  the multiplicative anomaly of determinants.  Differentiable families of symbols and operators are defined  in
the same way as were holomorphic families in Definitions
\ref{defn:holosymb} and \ref{defn:holomap} replacing holomorphic in the
parameter $z$  by
differentiable in the parameter $t$.
\begin{prop}\label{prop:difftrQ}
Let $A_t$ be  a differentiable family of $\Cl(M, E)$ of constant order $a$.
\begin{enumerate}
\item The noncommutative residue commutes with differentiation \begin{equation}\label{eq:resderiv}
\frac{d} {dt}{\rm res} (A_t)={\rm res} (\dot A_t),
\end{equation}
where we have set $\dot A_t=\frac{d}{dt} A_t$.
\item If the order $a$ is non integer, the canonical trace commutes with differentiation \begin{equation}\label{eq:TRderiv}
\frac{d} {dt}{\rm TR} (A_t)={\rm TR} (\dot A_t).
\end{equation}
 \item For any weight  $Q$
with  order $q$ and spectral cut $\alpha,$
\begin{equation}\label{eq:trQderiv}
\frac{d} {dt}{\rm tr}_{\alpha}^Q (A_t)={\rm tr}_{\alpha}^Q (\dot A_t).
\end{equation}
\end{enumerate}
\end{prop}
{\bf Proof:} Using (\ref{eq:sigmaN}) we write  the symbol $\sigma_{A_t}$ of
$A_t$ as follows:  $$\sigma_{A_t}(x, \xi)= \sum_{j=0}^{N-1}
 \chi(\xi)\,
 \left(\sigma_{A_t}\right)_{a-j}(x,\xi)+\left(\sigma_{A_t}\right)_{(N)}(x,
 \xi).$$
 \begin{enumerate}
\item By assumption,  the map $t\mapsto {\rm tr}_x\left( \left(\sigma_{A_t}(x,
    \cdot)\right)_{-n}\right)$ is
differentiable leading to a  differentiable map $t\mapsto \int_{S_x^*M}{\rm
  tr}_x\left(\left(\sigma_{A_t}(x, \cdot)\right)_{-n} \right)$  after integration over the compact set
$S_x^*M$ with  derivative:
 $t\mapsto \int_{S_x^*M}{\rm tr}_x\left(\dot \sigma_{A_t}\right)_{-n}$,  where $\dot
\sigma_{A_t}=\sigma_{\dot A_t}$ stands for the derivative of $\sigma_{A_t}$ at  $t$. Thus, the map
$t\mapsto {\rm res}(A_t)$ is differentiable with derivative given by
(\ref{eq:resderiv}).
\item By (\ref{eq:omegaKV}) and (\ref{eq:TRKV}), to prove formula
  (\ref{eq:TRderiv}) we need to check the differentiability of    the  map $t\mapsto \cutoffint_{T_x^*M}
{\rm tr}_x\sigma_{A_t}(x, \cdot)$ and to prove that
$$\frac{d}{dt}\cutoffint_{T_x^*M}
{\rm tr}_x\sigma_{A_t}(x, \cdot)=\cutoffint_{T_x^*M}
{\rm tr}_x\dot \sigma_{A_t}(x, \cdot) .$$ \\
The cut-off integral  involves  the whole symbol which we denote by $\sigma_t:=
\sigma_{A_t}$ in
order to simplify notations. Since the family $\sigma_t$ has constant order, $N$ can be  chosen independently of
$t$ in the asymptotic expansion. The corresponding cut-off integral  can be
computed explicitely (see e.g \cite{PS}):
\begin{eqnarray*}
\cutoffint_{T_x^*M}{\rm tr}_x(\sigma_t(x,\xi))\, \dbar\xi
 &=& \int_{T_x^*M}{\rm tr}_x\left(\left(\sigma_t\right)_{(N)}(x, \xi)\right) \, \dbar\xi+\sum_{j=0}^{N-1}
 \int_{\vert \xi\vert \leq 1}
 \chi(\xi)\,{\rm tr}_x\left(\left( \sigma_t\right)_{a-j}(x,\xi)\right) \, \dbar\xi\\
&- &\sum_{j=0, a-j+n\neq 0}^{N-1}  \frac{ 1}{a-j+n}
 \int_{\vert \xi\vert=1} {\rm tr}_x\left(\left(\sigma_t\right)_{a-j } (x,\omega)\right) \,\dbar_S\omega.
\end{eqnarray*}
 The map   $t\mapsto \int_{T_x^*M}{\rm tr}_x\left(
\left(\sigma_t\right)_{(N)}(x, \xi)\right) \,
\dbar\xi$  is differentiable at any point $t_0$ since by assumption  the maps $t\mapsto {\rm tr}_x\left( \left(\sigma_t(x,\xi)\right)_{(N)}\right)$ are
differentiable with modulus bounded from above $\left\vert  {\rm tr}_x\left(\left(\dot
\sigma_t(x,\xi)\right)_{(N)}\right)\right\vert \leq C\vert \xi\vert^{{\rm
    Re}(a)-N}$  by an $L^1$ function provided $N$ is chosen large enough, where the constant $C$ can be chosen independently of
     $t$ in a compact neighborhood of $t_0$.  Its
derivative is given by $t\mapsto \int_{T_x^*M}{\rm tr}_x\left(\left(\dot \sigma_t\right)_{(N)}(x, \xi)\right) \,
\dbar\xi$.  The remaining  integrals $ \int_{\vert \xi\vert \leq 1}
 \chi(\xi)\,{\rm tr}_x\left( \left(\sigma_t\right)_{a-j}(x,\xi)\right) \, \dbar\xi$ and $ \int_{\vert \xi\vert=1}
{\rm tr}_x\left( \left(\sigma_t\right)_{a-j } (x,\omega)\right) \,\dbar_S\omega$ are also differentiable
as integrals  over compact sets of  integrands involving differentiable maps
$t\mapsto{\rm tr}_x\left( \left(\sigma_t(x, \xi)\right)_{a-j}\right)$. Their
derivatives are given by  $ \int_{\vert \xi\vert \leq 1}
 \chi(\xi)\,{\rm tr}_x\left( \left(\dot \sigma_t\right)_{a-j}(x,\xi)\right) \, \dbar\xi$ and $ \int_{\vert \xi\vert=1}
 {\rm tr}_x\left(\left(\dot\sigma_t\right)_{a-j } (x,\omega)
 \right)\,\dbar_S\omega$. Thus,
$t\mapsto \cutoffint_{T_x^*M}{\rm tr}_x(\sigma_{A_t}(x,\xi))\, \dbar\xi$ is
differentiable with derivative given by $\cutoffint_{T_x^*M}{\rm
  tr}_x(\dot\sigma_{ A_t}(x,\xi))\, \dbar\xi$.
\item
By the defect formula (\ref{eq:trQPS}) we have
$$
{\rm tr}_\alpha^Q(A_t)= \int_M\, dx\left(\cutoffint_{T_x^*M} {\rm
    tr}_x\sigma_{A_t}(x, \cdot)-\frac{1}{q} \, \int_{S_x^*M}{\rm tr}_x\left(\sigma_{A_t\,
  \log_\alpha Q}(x,\cdot) \right)_{-n}\right)$$
which reduces the proof of the differentiability of $t\mapsto {\rm
  tr}_\alpha^Q(A_t)$ to that of the two maps $t\mapsto \cutoffint_{T_x^*M}
{\rm tr}_x\sigma_{A_t}(x, \cdot)$ and $t\mapsto \int_{S_x^*M}{\rm tr}_x\left(\sigma_{A_t\,
  \log_\alpha Q}(x, \xi)\right)_{-n}$.\\ \indent Differentiability of the first map was shown
in the second item of the proof. Let us first investigate the second map. By (\ref{eq:starproduct})
 we have  $$\left(\sigma_{A_t\,
  \log_\alpha Q}\right)_{-n}= \sum_{\vert \alpha\vert +a-j-k=-n }\frac{(-i)^{\vert
  \alpha\vert}}{\alpha!}\partial_{\xi}^{\alpha} \left(\sigma_{A_t}\right)_{a-j}\,
\partial_x^{\alpha} \left(\sigma_{\log_\alpha Q}\right)_{-k}.$$
By assumption,  the maps $t\mapsto \left(\sigma_{A_t}\right)_{a-j}$ are
differentiable so that  $t\mapsto \int_{S_x^*M}{\rm tr}_x\left(\sigma_{A_t\,
  \log_\alpha Q}\right)_{-n}$ is differentiable with derivative
 $$t\mapsto \int_{S_x^*M}{\rm tr}_x\left(\dot \sigma_{A_t\,
  \log_\alpha Q}\right)_{-n}= \int_{S_x^*M}{\rm tr}_x\left( \sigma_{\dot A_t\,
  \log_\alpha Q}\right)_{-n}. $$
 Integrating over the compact manifold $M$ then yields
that the map $t\mapsto {\rm tr}_{\alpha}^Q(A_t)$ is differentiable with
derivative given   by $$
 \int_M\, dx\left(\cutoffint_{T_x^*M} {\rm
    tr}_x\sigma_{\dot A_t}(x, \cdot)-\frac{1}{q} \, \int_{S_x^*M}{\rm
    tr}_x\left(\sigma_{\dot A_t\,
  \log_\alpha Q}(x,\cdot) \right)\right)={\rm tr}_\alpha^Q(\dot A_t).$$
\end{enumerate}
\endsquare
\section{Locality of weighted traces of $L(A,B)$}
 Combining (\ref{eq:Lbracket}) with (\ref{eq:coboundary}) yields the locality
 of weighted traces ${\rm tr}^Q(L(A,B))$ as a finite sum of noncommutative
   residues, independently of the choice of spectral cut. \\
\indent In this section we  show
 that  weighted traces of $L(A,B)$  only depend
 on a finite number of homogeneous components of the operators $A$ and $B$
 (see Theorem \ref{thm:trQLABlocal}), a fact reminiscent of a similar property
 observed by Okikiolu in \cite{O1} in the case of operators with scalar
 leading symbols.\\
\begin{lem}\label{lem:dlogA} Let $A_t$ be a differentiable  family of
  admissible operators in $\Cl(M, E)$ with constant spectral cut $\alpha$. For any positive integer $K$ we have
$$\frac{d}{dt} \log_\alpha A_t= \dot A_t A_t^{-1}+ \sum_{k=1}^K {\rm ad}^{k}_{A_t} (\dot
A_t) \, A_t^{-(k+1)}+ R_K(A_t, \dot A_t)$$ where we have set $\dot
A_t:= \frac{d}{dt}A_t$ and \begin{eqnarray}\label{eq:RK}R_K(A_t, \dot A_t)&:=&\frac{d}{dz}\left(
\frac{i}{2\pi} \int_{\Gamma_\alpha} \lambda^z\,\left[(\lambda-A_t)^{-1}, {\rm
ad}_{A_t}^K (\dot A_t)\right]
(\lambda-A_t)^{-(K+1)}\,d\lambda\right)_{\vert_{z=0}}\\
&=&{\rm
ad}_{A_t}^K \left( \frac{d}{dz}\left(
\frac{i}{2\pi} \int_{\Gamma_\alpha} \lambda^z\,\left[(\lambda-A_t)^{-1}, \dot A_t\right]
(\lambda-A_t)^{-(K+1)}\,d\lambda\right)_{\vert_{z=0}}\right)\nonumber,
\end{eqnarray} 
since $A_t$ commutes with $(\lambda-A_t)^{-k}$. \\ Here $\Gamma_\alpha$ is a
contour around the spectrum as in (\ref{eq:contourGamma}). \\ \indent
If $A_t$ commutes with $\dot A_t$ then $\frac{d}{dt} \log_\alpha A_t= \dot
A_t A_t^{-1}$.  If  $A_t$ has constant order $a$ then  $\frac{d}{dt} \log A_t$ lies in $\Cl^0(M, E)$.
\end{lem}
{\bf Proof:} Since the spectral cut is constant, we drop it in the notation
setting $\log A_t= \log_\alpha A_t$. \\ \indent
Since $\log A_t$ has a symbol of the form $\sigma_t\sim a_t\, \log \vert \xi\vert
+\sigma_{0,t}(x, \xi)$ where $a_t$ is the  order of $A_t$ and
$\sigma_{0,t}$ a symbol of order $0$, if $a_t=a$ is constant, its derivative $\frac{d}{dt} \log A_t$,
 has a symbol of the form
 $\frac{d}{dt} \sigma_t\sim
\frac{d}{dt}  \sigma_{0,t}(x, \xi)$ and  is therefore classical of zero order. \\ \indent
Going back to the general situation and assuming differentiability of the family at $0$, we  derive the
formula for the derivative at $0$; a change of variable $t\to t-t_0$ then
yields the corresponding result at any point $t_0$ of differentiability. We set  $\Delta_t:= \frac{A_t-A_0}{t}$ which tends to
$\dot A_0$ as $t$ tends to zero.  We observe that
$
(\lambda-A_0)( \lambda-A_t)^{-1}= (\lambda-A_t)( \lambda-A_t)^{-1}
+ t\, \Delta_t ( \lambda-A_t)^{-1}= 1+ t\, \Delta_t( \lambda-A_t)^{-1}$,
 from which we infer that $$( \lambda-A_t)^{-1}= (\lambda-A_0)^{-1}+
t\, (\lambda-A_0)^{-1}\Delta_t ( \lambda-A_t)^{-1}.$$ By  induction
we get
$$( \lambda-A_t)^{-1}=(\lambda-A_0)^{-1}+\sum_{p=1}^P t^p\,
(\lambda-A_0)^{-1}\Delta_t\cdots \Delta_t(\lambda-A_0)^{-1}\Delta_t
(\lambda-A_0)^{-1}+ t^{P+1}\, R_P(A_0, \Delta_t,\lambda),$$ where $ R_P(A_0, \Delta_t,\lambda) $ stands for the remainder
term. It follows that $$\lim_{t\to 0}\frac{(A_t-\lambda)^{-1}-
  (A_0-\lambda)^{-1}}{t}=(\lambda-A_0)^{-1}\,\lim_{t\to 0}\Delta_t\,
(\lambda-A_0)^{-1}=(\lambda-A_0)^{-1} \dot A_0 (\lambda-A_0)^{-1}.$$
For $t$ in a compact neighborhood of $0$, we can choose a common  contour
$\Gamma$ along a spectral ray around the spectrum  of $A_t$ as in
(\ref{eq:contourGamma}). Hence,
\begin{eqnarray*}
 \frac{d}{dt}_{\vert_{t=0}}\log(A_t)&= &\lim_{t\to 0}\frac{\log(A_t)-\log (A_0)}{t}\\
&=&  \frac{d}{dz}\left(\frac{i}{2\pi} \int_\Gamma \lambda^z \frac{
d}{dt}_{\vert_{t=0}}\left((A_t-\lambda)^{-1}-
  (A_0-\lambda)^{-1}\right)\, d\lambda\right)_{\vert_{z=0}}\\
&=& \frac{d}{dz}\left(\frac{i}{2\pi} \int_\Gamma \lambda^z
\,(\lambda-A_0)^{-1}\dot A_0 (\lambda-A_0)^{-1}\,
d\lambda\right)_{\vert_{z=0}}.
\end{eqnarray*}
On the other hand, \begin{eqnarray*}
\left[(\lambda-A_0)^{-1}, \dot A_0\right]&=& (\lambda-A_0)^{-1}\, [A_0, \dot A_0]\,
(\lambda-A_0)^{-1}\\
&=& [A_0, \dot A_0] (\lambda-A_0)^{-2}+  {\rm ad}_{A_0}^2( \dot A_0)\,
(\lambda-A_0)^{-3}+\left[(\lambda-A_0)^{-1}, {\rm ad}_{A_0}^2( \dot A_0)\right]
(\lambda-A_0)^{-2}\\
&=&\sum_{k=1}^K {\rm ad}_{A_0}^k( \dot A_0)
(\lambda-A_0)^{-(k+1)}+  \left[(\lambda-A_0)^{-1},{\rm ad}_{A_0}^{K}(\dot A_0)\right]
(\lambda-A_0)^{-K},
\end{eqnarray*}
Hence,
\begin{eqnarray*}
\frac{d}{dt}_{ \vert_{t=0}}\log
(A_t)&=&\frac{d}{dz}\left(\frac{i}{2\pi} \int_\Gamma \lambda^z
\,(\lambda-A_0)^{-1}\dot A_0(\lambda-A_0)^{-1}\,
d\lambda\right)_{\vert_{z=0}}\\
&=&\dot A_0\, \frac{d}{dz}\left(\frac{i}{2\pi} \int_\Gamma
\lambda^z \,(\lambda-A_0)^{-2}\, d\lambda\right)_{\vert_{z=0}}\\
&+& \sum_{k=1}^K {\rm ad}_{A_0}^{k}(\dot A_0)
\frac{d}{dz}\left(\frac{i}{2\pi} \int_\Gamma
\lambda^z \,(\lambda-A_0)^{-(k+2)}\,d\lambda\right)_{\vert_{z=0}}\\
&+& \frac{d}{dz}\left( \frac{i}{2\pi} \int_\Gamma
\lambda^z\, \left[(\lambda-A_0)^{-1},{\rm ad}_{A_0}^{K}(\dot A_0)\right]
(\lambda-A_0)^{-(K+1)}\,d\lambda\right)_{\vert_{z=0}}.
\end{eqnarray*}
Iterated integrations by parts yield
$$
\frac{d}{dt}_{\vert_{t=0}}\log (A_t) = \dot A_0\, A_0^{-1}+
\sum_{k=1}^K {\rm ad}_{A_0}^{k}(\dot A_0) A_0^{-(k+1)}+ R_K(A_0,
\dot A_0),$$ where  $R_K(A_0,\dot A_0)=\frac{d}{dz}\left(
\frac{i}{2\pi} \int_\Gamma \lambda^z\,\left[(\lambda-A_0)^{-1}, {\rm
ad}_{A_0}^{K}(\dot A_0)\right]
(\lambda-A_0)^{-(K+1)}\,d\lambda\right)_{\vert_{z=0}}$ is the remainder
term.\endsquare\\\\\indent
The following result is reminiscent of an observation made in
\cite{O1} (see also \cite{Sc}), namely  that only the first $n$
homogeneous components of the symbols come into play for the
derivation of the Campbell-Hausdorff formula for operators with
scalar leading symbols; the weighted trace of $L(A,B)$ presents  a
similar feature in our more general situation.
\begin{thm}\label{thm:trQLABlocal}
Given a weight $Q $  and two admissible operators $A$ and $B$ in $\Cl(M, E)$
with non negative orders,
the weighted trace ${\rm tr}^Q(L(A,B))$ is a local expression as a finite sum
of noncommutative
residues, which  only depends on the first $n$
homogeneous components of the symbols of $A$ and $B$:
\begin{equation}\label{eq:partialtvanishes}\frac{ d}{dt} {\rm tr}^Q(L(A(1+tS), B)= \frac{ d}{dt} {\rm tr}^Q(L(A, B(1+tS))=
0\quad\forall S\in \Cl^{<-n}(M, E),\end{equation} where $\Cl^{<-n}(M,
E)=\cup_{{\rm Re}(a)<-n}\Cl^a(M, E)$
stands for the algebra of classical operators of order with real part $<-n$.
\end{thm}
{\bf Proof:}\begin{itemize}
\item
On the one hand we know that  $L(A,B)$ is a finite sum of commutators
of  classical pseudodifferential operators $[P_j, Q_j]$. By
(\ref{eq:coboundary}),
each weighted trace ${\rm tr}^Q([P_j, Q_j]$ is proportional to ${\rm res} \left( Q_j \, [P_j,
  \log_\alpha Q]\right)$ so that  ${\rm tr}^Q(L(A,B))$ is  indeed a finite sum
of noncommutative
residues.
\item  Let us  check that requirement (\ref{eq:partialtvanishes})
is equivalent to the fact that  ${\rm tr}^Q(L(A,B))$  only depends on the first $n$
homogeneous components of the symbols of $A$ and $B$. \\ \indent
Given an operator $S$  in $\Cl(M, E)$ of
order $<-n$ and an operator $A$  in $\Cl(M, E)$ of order $a$, we first observe that   in any local trivialisation the first $n$ homogeneous components of the symbols of
 $A$ and $A(1+S)$ coincide since $AS$ has order $a-n$.  Conversely, if  the
 first $n$ homogeneous components of the symbols of two  classical operators
 $A$ and $B$ of orders $a$ and $b$ coincide, then  $a=b$ and if $B$ is
 invertible,  the first $n$
 homogeneous components of the symbol of  $B^{-1}$  defined inductively using
 (\ref{eq:starproduct})  by:
$$\left(\sigma_{B^{-1}}\right)_{-b}=\left(\left(\sigma_B\right)_{b}\right)^{-1},$$
$$\left(\sigma_{B^{-1}}\right)_{-b-j}=-\left(\left(\sigma_B\right)_{b}\right)^{-1}\,
\sum_{k+l+\vert \alpha\vert=j,l<j} \frac{(-i)^{\vert
    \alpha\vert}}{\alpha!}\partial_\xi^\alpha \left(\sigma_B\right)_{b-k} \,
\partial_x ^\alpha  \left(\sigma_{B^{-1}}\right)_{-b-l}, $$
coincide with that of the symbol of $A^{-1}$ since the terms
corresponding to $j\leq n$ only involve homogeneous components
$\left(\sigma_B\right)_{b-k}=\left(\sigma_A\right)_{a-k} $ and
$\left(\sigma_{B^{-1}}\right)_{-b-l}$ with $k$ and $l$ no larger
than $n$. Consequently, by  (\ref{eq:starproduct}) it follows that
$S= A^{-1}\, B$ has order $<-n$.  Thus, showing that the expression
${\rm tr}^Q(L(A,B))$ only depends on the first $n$ homogeneous
components of $A$ amounts to showing that  ${\rm tr}^Q(L(A+S,B))=
{\rm tr}^Q(L(A,B))$ for any classical operator $S$ of order $<-n$.
  \item Let us further observe that the proof of  (\ref{eq:partialtvanishes})  reduces to the proof at
    $t_0=0$. Indeed  for any real number $t_0$, we have
 \begin{eqnarray}\label{eq:implication}
\frac{ d}{dt}_{\vert_{t=0}} {\rm tr}^Q(L(C(1+tT), D)&=&0\quad\forall T\in
 \Cl^{<-n}(M, E),\quad \forall \quad {\rm admissible}\quad  C, D \quad
 \nonumber\\
\Longrightarrow
 \frac{ d}{dt}_{\vert_{t=t_0}} {\rm tr}^Q(L(A(1+tS), B)&=&0\quad\quad\forall S\in
 \Cl^{<-n}(M, E),\quad \forall\quad  {\rm admissible}\quad  A, B
\nonumber.
\end{eqnarray}
To check  this implication, we set
$u=t-t_0$ so that
$1+tS= 1+t_0S+uS= (1+uS(1+t_0S)^{-1})(1+t_0S)$. Setting $T= S(1+t_0S)^{-1}$
which also lies in $ \Cl^{<-n}(M, E)$, we have
\begin{eqnarray*}
&&L(A(1+tS), B)- L(A(1+u T), (1+t_0S)B)\\
&=& -\log (A(1+uT)(1+t_0S))-\log (B)+\log A(1+uT)+\log ((1+t_0S)B)\\
&=&
 L(1+t_0S, B)-L(A(1+uT),1+t_0S),
\end{eqnarray*}
and hence
$$L(A(1+tS), B)= L(A(1+u T), (1+t_0S)B)+L(1+t_0S, B)-L(A(1+uT),1+t_0S).$$
Differentiating w.r. to $t$ at $t=t_0$ on the l.h.s boils down to
differentiating the r.h.s. at $u=0$  and  the  implication
(\ref{eq:implication}) then easily
follows.
 \item We are therefore left to prove that $ \frac{ d}{dt}_{\vert_{t=0}} {\rm tr}^Q(L(A(1+tS), B)= 0$.
Applying (\ref{eq:trQderiv}) to the operator $A_t:=L(A(1+tS), B)$ we have
$$\frac{ d}{dt}_{\vert_{t=0}} {\rm tr}^Q(L(A(1+tS), B)= {\rm tr}^Q \left(\frac{ d}{dt}_{\vert_{t=0}}
  \left(L(A(1+tS),B)\right)\right).$$ We therefore need to investigate
the behaviour of $\frac{L(A(1+tS),B)-L(A,B)}{t}$ as $t\to 0$.
Since
$$L(A(1+tS),B)-L(A,B)= \log (A(1+tS)B)-\log (AB) -\left(\log(A(1+tS))-\log
  A\right),$$
let us study  the difference
 $\log (A(1+tS)C)-\log (AC)$ with $C$ equal to either $B$ or the identity
 operator.
Let us apply  Lemma \ref{lem:dlogA} to  $A_t:=A(1+tS)C$ so that $A_0=AC$. When
$t$ varies in a small compact neighborhood of $0$, the operators $A_t$ have a common
spectral cut $\alpha$ which we drop in the notation.  Implementing  the weighted trace ${\rm tr}^Q$ yields
\begin{eqnarray*}
&&\frac{d}{dt}_{\vert_{t=0}}{\rm tr}^Q\left(\log (A(1+tS)C)\right)\\ 
&=&{\rm tr}^Q( ASC\, (AC)^{-1})+ \sum_{k=1}^K {\rm tr}^Q({\rm ad}_{AC}^k(ASC)\,(AC)^{-(k+1)})+
{\rm tr}^Q(R_K(AC, ASC))
\end{eqnarray*}
for arbitrary large $K$ and with remainder term
 $$R_K(AC, ASC):= \frac{d}{dz}\left(
\frac{i}{2\pi} \int_{\Gamma_\alpha} \lambda^z\,\left[(\lambda-AC)^{-1}, {\rm
ad}_{AC}^K \left( ASC\right)\right]
(\lambda-AC)^{-(K+1)}\,d\lambda\right)_{\vert_{z=0}}.$$ \indent 
But for any positive integer $k$, by
(\ref{eq:coboundary}) we have
\begin{eqnarray*}
{\rm tr}^Q({\rm ad}_{AC}^k(ASC)\, (AC)^{-(k+1)}))&= &{\rm tr}^Q\left({\rm ad}_{AC}
({\rm ad}_{AC}^{k-1}(ASC))\,(AC)^{-(k+1)}\right)\\
&= & {\rm tr}^Q\left({\rm ad}_{AC}
\left({\rm ad}_{AC}^{k-1}(ASC) \, (AC)^{-(k+1)}\right)\right)\\
&=&-\frac{1}{q} {\rm res}\left({\rm ad}_{AC}^{k-1}(ASC)\, (AC)^{-(k+1)}\, [AC,
\log Q]\right)\\ &=& 0.
\end{eqnarray*}
Here we use the fact that the operator ${\rm ad}_{AC}^{k-1}(ASC)\, (AC)^{-(k+1)}\, [AC,
\log Q] $ has order $(k-1)(a+c)+a+c+s-(k+1)(a+c)+a+c=s$ (here $s$ is the order
of $S$, $a$ the order of $A$, $c$ the order of $C$)
with   real part  smaller than $-n$. Thus
$$\frac{d}{dt}_{\vert_{t=0}}{\rm tr}^Q\left(
\log (A(1+tS)C)\right)= {\rm tr}^Q\left(ASC\,    \, (AC)^{-1}\right)
+ {\rm tr}^Q\left(R_K(AC, ASC)\right),$$ independently of the choice
of the integer $K$. The remainder term ${\rm tr}^Q(R_K(AC, ASC))$
depends on $S$ via the iterated brackets ${\rm
  ad}_{AC}^{K}(ASC)$ and hence via $K$. Since it is independent of $K$,  it is
also be independent of $S$. Setting $S=0$ which lies in $\Cl^{<-n}(M, E)$, we
infer that ${\rm tr}^Q(R_K(AC, ASC))$ vanishes for all positive integers
$K$. Thus $$\frac{d}{dt}_{\vert_{t=0}}{\rm tr}^Q\left(
\log (A(1+tS)C)\right)= {\rm tr}^Q\left(ASC\,    \, (AC)^{-1}\right) = {\rm
tr}^Q(ASA^{-1}),$$ independently of $C$.
Setting back $C=B$ and $C=I$ yields
 \begin{eqnarray*}
 \frac{ d}{dt}_{\vert_{t=0}}  {\rm tr}^Q(L(A(1+tS), B))&=&  {\rm tr}^Q\left(\frac{ d}{dt}_{\vert_{t=0}} \log
 (A(1+tS)B\right)-   {\rm tr}^Q\left(\frac{ d}{dt}_{\vert_{t=0}} \log
 (A(1+tS)\right)\\
&=&0
\end{eqnarray*}
thus ending  the proof of the Theorem.
\end{itemize}
\endsquare

\section{A local formula for the  weighted trace of $L(A,B)$ }
\indent We  derive  an explicit  local
expression for the weighted traces ${\rm tr}^A(L(A,B))$ and ${\rm
  tr}^B(L(A,B))$ of $L(A,B)$ (see Theorem \ref{thm:weightedTraceLAB}). Our
approach is  inspired by the Okikiolu's proof  for the Campbell-Hausdorff formula for operators with
scalar leading symbols. In the case of operators with scalar leading
symbols, as it was noticed and used by Okikiolu, as from a certain
order in the Campbell-Hausdorff expansion,  one can implement
ordinary traces since the iterated brackets have decreasing order.
In our more general situation, such a phenomenon does not occur so
that  we  use weighted traces instead.
\begin{prop}\label{prop:dercanTraceLAB}
 Let $A$ and $ B$   be two admissible  operators with  positive  orders
 $a$ and $
 b$ in $\Cl(M, E)$ such that their
  product $AB$ is also admissible.  We have the following  identities for  weighted traces:
$${\frac{d}{dt}}_{\vert {t=0}} {\rm
tr}^B(L(A^t,B^\mu))=0,\quad\quad{\frac{d}{dt}}_{\vert {t=0}} {\rm
tr}^A(L(A^t,B^\mu))=0 $$ as well as for the
noncommutative residue:
$${\frac{d}{dt}}_{\vert {t=0}} {\rm
res}(L(A^t,B^\mu))=0.$$
\end{prop}
{\bf Proof:} Let us  prove the result for the
$B$-weighted trace; a similar proof yields the result for the $A$-weighted
trace. By Proposition \ref{prop:difftrQ},  weighted traces and the residue commute with
differentiation on constant order operator so that
$${\frac{d}{dt}}_{\vert_{t=0}}{\rm tr}^Q\left(L(A^t,B^\mu)\right)=
{\rm tr}^Q\left({\frac{d}{dt}}_{\vert_{t=0}}L(A^t,B^\mu)\,\right)$$ resp.
$$ \quad  {\frac{d}{dt}}_{\vert {t=0}}{\rm res}\left(L(A^t,B^\mu)\right)=
{\rm res}\left({\frac{d}{dt}}_{\vert {t=0}}(L(A^t,B^\mu)\,
\right).$$
But
$$ {\frac{d}{dt}}_{\vert {t=0}}L(A^t,B^\mu)= {\frac{d}{dt}}_{\vert {t=0}}\log
(A^tB^\mu)-{\frac{d}{dt}}_{\vert {t=0}}\log A^t.$$
We therefore apply  Lemma \ref{lem:dlogA} to  $A_t:=A^tB^\mu$ so that
$A_0=B^\mu$, including the case $\mu=0$  for which $A_t=A^t$ and $A_0=I$.
Since $\dot A_0= \log A\, B^\mu$ and
$\dot A_0 \, A_0^{-1}= \log A$, implementing  the weighted trace ${\rm tr}^B$ yields
\begin{eqnarray*}
&&\frac{d}{dt}_{\vert_{t=0}}{\rm tr}^B\left(\log (A^tB^\mu)\right)\\
&=&{\rm tr}^B(\log  A)+ \sum_{k=1}^K
{\rm tr}^B({\rm ad}_{B^\mu}^k(\log  A\,B^\mu)\,B^{-\mu(k+1)})+
{\rm tr}^B(R_K(B^\mu, \log  A\,B^\mu))
\end{eqnarray*}
for arbitrary large $K$, with  remainder term 
\begin{eqnarray*}
R_K(B^\mu, \log  A\,B^\mu)&= & \frac{d}{dz}\left(
\frac{i}{2\pi} \int_{\Gamma_\alpha} \lambda^z\,\left[(\lambda-B^\mu)^{-1}, {\rm
ad}_{B^\mu}^K(\log A\, B^\mu)\right]
(\lambda-B^\mu)^{-(K+1)}\,d\lambda\right)_{\vert_{z=0}}\\
&=& {\rm
ad}_{B^\mu}^K \left( \frac{d}{dz}\left(
\frac{i}{2\pi} \int_{\Gamma_\alpha} \lambda^z\, \left[(\lambda-B^\mu)^{-1}, log A\, B^\mu\right]
(\lambda-B^\mu)^{-(K+1)}\,d\lambda\right)_{\vert_{z=0}} \right),
\end{eqnarray*}
since $B$ commutes with $ B^\mu$.\\
 For any positive integer $k$, by
(\ref{eq:coboundary}) we have
\begin{eqnarray*}
{\rm tr}^B({\rm ad}_{B^\mu}^k(A\,B^\mu)\, B^{-\mu(k+1)}))&= &{\rm tr}^B\left({\rm ad}_{B^\mu}
({\rm ad}_{B^\mu}^{k-1}(A\,B^\mu))\,B^{-\mu(k+1)})\right)\\
&= & {\rm tr}^B\left({\rm ad}_{B^\mu}
\left({\rm ad}_{B^\mu}^{k-1}(A\, B^\mu) \, B^{-\mu (k+1)}\right)\right)\\
&=&-\frac{1}{b} {\rm res}\left({\rm ad}_{B^\mu}^{k-1}(A\, B^\mu)\, B^{-\mu(k+1)}\, [B^\mu,
\log B]\right)\\ &=& 0,
\end{eqnarray*}
since $B$ commutes with $\log B$. A similar computation shows that ${\rm
  tr}^B(R_K(B^\mu, \log  A\,B^\mu))=0$.  Thus
$$\frac{d}{dt}_{\vert_{t=0}}{\rm tr}^B\left(
\log (A^tB^\mu)\right)= {\rm tr}^B\left(\log A\right).$$
 It follows that $\frac{d}{dt}_{\vert_{t=0}}{\rm tr}^B\left(
\log (A^tB^\mu)\right)= {\rm tr}^B\left(\log A\right)$ independently of $\mu$
so that
$${\frac{d}{dt}}_{\vert {t=0}}{\rm tr}^B\left(L(A^t,B^\mu)\right)=0.$$
Similarly, replacing the weighted trace ${\rm tr}^B$ by the noncommutative residue ${\rm
  res}$ and using the cyclicity of the noncommutative residue, yields
$${\frac{d}{dt}}_{\vert {t=0}}{\rm res}\left(L(A^t,B^\mu)\right)=0.$$\endsquare\\
\indent The following statement  provides a local formula for the
  multiplicative anomaly of the zeta determinant. It also  shows that the
  residue of $L(A,B)$ vanishes and therefore yields back  the multiplicativity of the residue
  determinant derived in \cite{Sc}.
\begin{thm}\label{thm:weightedTraceLAB}
 For two admissible  operators $A, B\in \Cl(M, E)$    with positive orders
 $a$ and  $b$ such that their
  product $AB$ is also admissible, we have
\begin{equation}\label{eq:resLvanishes}{\rm res}(L(A,B))=0.\end{equation}
Moreover, there is an operator \begin{equation}\label{eq:Wtau}
    W(\tau)(A,B):={\frac{d}{dt}}_{\vert {t=0}} L(A^t, A^\tau B)\end{equation}
 in $\Cl^0(M, E)$   depending continuously on $\tau$ such that
\begin{equation}\label{eq:trQL}
{\rm tr}^Q(L(A,B))= \int_0^1{\rm
  res}\left(W(\tau)(A,B)\left(\frac{\log(A^\tau B)}{a\tau+b}
-\frac{\log Q}{q}\right)\right)d\tau
\end{equation}
where $Q$ is any weight of order $q$.
\end{thm}
{\bf Proof:} By  Proposition \ref{prop:dercanTraceLAB},  we know
that ${\frac{d}{dt}}_{\vert {t=0}} {\rm
res}(L(A^t,B))={\frac{d}{dt}}_{\vert {t=0}} {\rm tr}^Q(L(A^t,B))=0.$
We want to compute${\frac{d}{dt}}_{\vert {t=\tau}} {\rm
    res}(L(A^t,B))={\frac{d}{dt}}_{\vert {t=0}} {\rm
    res}(L(A^{t+\tau},B))$  and   ${\frac{d}{dt}}_{\vert {t=\tau}} {\rm
    tr}^Q(L(A^t,B))={\frac{d}{dt}}_{\vert {t=0}} {\rm
    tr}^Q(L(A^{t+\tau},B)).$ For this we observe that
$$L(AB,D)- L(A, BD)
= -\log (AB)-\log (D)+\log A+\log (BD)= L(B, D)-L(A,B)$$
Replacing $A$ by $A^t$, $B$ by $A^\tau$ and $D$ by $B$, we get
$$L(A^{t+\tau},B)- L(A^t, A^\tau B)
=  L(A^\tau, B)-L(A^t,A^\tau)=L(A^\tau, B).$$
 Implementing
the noncommutative residue, by Proposition \ref{prop:dercanTraceLAB} we have:
\begin{eqnarray*}
{\frac{d}{dt}}_{\vert {t=\tau}} {\rm res} (L(A^t,B))
&=&{\frac{d}{dt}}_{\vert {t=0}} {\rm res}(L(A^{t+\tau},B))\\
&=& {\frac{d}{dt}}_{\vert {t=0}}{\rm res}( L(A^t, A^\tau B))\\
&=&0.
\end{eqnarray*}
Hence \begin{equation}\label{eq:resL}
{\rm res}(L(A,B))= \int_0^1{\frac{d}{dt}}_{\vert {t=\tau}} {\rm
  res}(L(A^t,B))\, d\tau+ {\rm res}(L(I, B))=0,
\end{equation}
since $L(I,B)=0$. \\
If instead we implement the weighted trace ${\rm tr}^Q$, we have:
\begin{eqnarray*}
{\frac{d}{dt}}_{\vert {t=\tau}} {\rm tr}^Q (L(A^t,B))
&=&{\frac{d}{dt}}_{\vert {t=0}} {\rm tr}^Q(L(A^{t+\tau},B))\\
&=& {\frac{d}{dt}}_{\vert {t=0}}{\rm tr}^Q( L(A^t, A^\tau B)).
\end{eqnarray*}
Since  $A$ and $B$ have positive order so has $A^\tau \, B$, so that applying  Proposition
\ref{prop:dercanTraceLAB} with weighted traces  $ {\rm tr}^{A^\tau \, B}$ yields:
\begin{eqnarray*}
{\frac{d}{dt}}_{\vert {t=\tau}} {\rm tr}^Q(L(A^t\, B)
&=&{\frac{d}{dt}}_{\vert {t=0}}{\rm tr}^Q( L(A^t, A^\tau B))\\
&=& {\frac{d}{dt}}_{\vert {t=0}}{\rm tr}^{A^\tau B}( L(A^t, A^\tau B))\\
&+& {\frac{d}{dt}}_{\vert {t=0}}\left({\rm tr}^Q( L(A^t, A^\tau B)) -
    {\rm tr}^{A^\tau B}( L(A^t, A^\tau B))\right)\\
&=& {\frac{d}{dt}}_{\vert {t=0}}\left({\rm tr}^Q( L(A^t, A^\tau B)) -
    {\rm tr}^{A^\tau B}( L(A^t, A^\tau B))\right).
\end{eqnarray*}
Applying (\ref{eq:differenceWeighTrace})  to $Q_1= Q$ and
$Q_2= A^\tau B$, we infer that
\begin{eqnarray*}&& {\frac{d}{dt}}_{\vert {t=0}}\left({\rm tr}^Q( L(A^t, A^\tau B)) -
    {\rm tr}^{A^\tau B}( L(A^t, A^\tau B))\right)\\
&=&
 {\frac{d}{dt}}_{\vert {t=0}}{\rm
  res}\left( L(A^t, A^\tau B)\left(\frac{\log(A^\tau B)}{a\tau+b}
-\frac{\log Q}{q}\right)\right)\\
&=&{\rm
  res}\left(W(\tau)(A,B)\,\left(\frac{\log(A^\tau B)}{a\tau+b}
-\frac{\log Q}{q}\right)\right),
 \end{eqnarray*}
where $q$ is the order of $Q$ and where we have set
$W(\tau)(A,B):={\frac{d}{dt}}_{\vert {t=0}} L(A^t, A^\tau B)$.
Since  $L(I,B)=0$, we finally find that
 \begin{eqnarray}\label{eq:weightedTraceLAB}
&{}&{\rm  tr}^Q(L(A,B))
={\rm  tr}^Q(L(A^1,B))-{\rm tr}^Q(L(A^0,B))\nonumber\\
&=& \int_0^1{\rm
  res}\left( W(\tau)(A,B)\,\left(\frac{\log(A^\tau B)}{a\tau+b}
-\frac{\log Q}{q}\right)\right) d\tau.
\end{eqnarray}
\endsquare

\section{Multiplicative anomaly for determinants revisited}
\setcounter{equation}{0} \indent We first observe that the  multiplicative
anomaly for weighted determinants studied in \cite{D}  has logarithm
given by the weighted trace of $L(A,B)$, as a result of which it is
local. We then derive an explicit local formula for the
multiplicative anomaly of $\zeta$ determinants, using the local
formula derived previously for
weighted traces of $L(A,B)$. \\
\\  An   admissible  operator $A\in
\Cl(M, E)$ with spectral cut $\theta$ and  positive order  has well
defined $Q$-weighted determinant \cite{D} (see also \cite{FrG})
where $Q\in \Cl(M, E)$ is a weight with spectral cut $\alpha$:
$${\rm det}_\alpha^Q(A):= e^{{\rm tr}_\alpha^Q(\log_\theta A)}.$$
Here the weighted trace has been extended to logarithms as before,
picking out the constant term of the meromorphic  map $z\mapsto {\rm
TR}(\log_\theta  A\, Q_\alpha^{-z})$ which can have double poles in contrast
to
 the case of classical operators studied in Section 3.
\begin{rk}{\rm  The weighted
determinant, as well as being dependent on the choice of spectral cut
$\theta$,  also  depends on the choice of spectral cut $\alpha$.}
\end{rk}\indent
Since the weighted trace restricts to the ordinary trace on
trace-class operators, this determinant, as the $\zeta$-determinant,
extends the ordinary determinant on operators in the determinant
class.
 \\
 \begin{lem}\label{lem:detQspectralcut} Let $0\leq\theta<\phi<2\pi$ be two
spectral cuts for
the admissible operator $A$.  If  there is a cone $\Lambda_{\theta,
\phi}$ (see \ref{eq:cone}) which does not intersect the
spectrum of the leading symbol of $A$ then
$${\rm det}_{ \theta}^Q(A)={\rm det}_{\phi}^Q(A).$$
\end{lem}
{\bf Proof:}  Under the assumptions of the proposition,the cone
$\Lambda_{\phi, \theta}$ defined as in Proposition
\ref{prop:spectralcutproj}, contains only a finite number of points
in the spectrum of $A$  so that  $\log_\phi A-\log_\theta
A=2i\pi\Pi_{\theta, \phi}(A)$ is a finite rank operator and hence
smoothing. Hence,

\begin{eqnarray*} \frac{{\rm det}^Q_{\phi}(A)}{{\rm
det}_{\theta}^Q(A)}& =& e^{{\rm tr}^Q\left(\log_\phi A-\log_\theta
A\right)}= e^{ {\rm tr}^Q\left(2i \pi\,\Pi_{\theta, \phi}(A)\right)}\\
& =& e^{2i \pi\, {\rm tr}\left(\Pi_{\theta, \phi}(A)\right)}= e^{2i\pi \, {\rm rk} (\Pi_{\theta,
  \phi}(A))} \\
  &=& 1,
\end{eqnarray*} where rk stands for the rank.
\endsquare\vskip 0,3cm \indent The
multiplicative anomaly for $Q$-weighted determinants  of two admissible operators
$A$, $B$ with spectral cuts $\theta, \phi$ such that $AB$ has
spectral cut $\psi$ is defined by:
$${\cal M}_{\theta, \phi, \psi}^Q(A, B):= \frac{{\rm det}_{\psi}^Q(AB)}{{\rm det}_{
\theta}^Q(A)\,{\rm
    det}_{ \phi}^Q(B)}, $$
    which we write ${\cal M}^Q(A, B)$ for simplicity.
\begin{prop}\label{prop:weightedmultanom} Let
$A$ and  $B$ be two admissible operators with spectral cuts $\theta$ and $\phi$  in $[0,2 \pi[$ such that
there is a cone delimited by the rays $L_\theta$ and $L_\phi$
which does not intersect the spectra of the leading symbols of $A$, $B$ and
$AB$. Then the product
$AB$  is admissible with a spectral cut $\psi$ inside that cone and for any
weight  $Q$
with spectral cut, dropping the explicit mention of the spectral cuts we have:
\begin{equation}\label{eq:detQanomaly}
\log {\cal M}^Q(A, B)= \int_0^1{\rm
  res}\left(W(\tau)(A,B)\left(\frac{\log(A^\tau B)}{a\tau+b}
-\frac{\log_\alpha Q}{q}\right)\right)d\tau.
\end{equation}
Weighted determinants are multiplicative on commuting operators.
\end{prop}
{\bf Proof:}
Since the leading symbol of the product $AB$ has spectrum which
does not intersect the cone delimited by $L_\theta$ and $L_\phi$, the operator
$AB$ only has a finite number of eigenvalues inside that cone. We can
therefore choose
a ray $\psi$ which avoids both the spectrum of the leading symbol of $AB$ and
the eigenvalues of $AB$. By the above lemma, the   weighted
determinants ${\rm det}_\theta^Q(A)$, ${\rm det}_\phi^Q(B)$ and
${\rm det}_\psi^Q(AB)$ do not depend on the choices of  spectral
cuts satisfying the requirements of the proposition.\\
 Since
$$ \log {\cal M}^Q(A, B) =\log {\rm det}^Q(AB)- \log {\rm det}^Q(A)-\log {\rm det}^Q(B)={\rm tr}^Q(L(A,B)),$$
the logarithm of the multiplicative anomaly for weighted determinants  is  a
local quantity  (\ref{eq:trQL}) derived in Theorem \ref{thm:weightedTraceLAB}.\\ To prove the second part of the statement we observe that
\begin{equation}\label{eq:commutingLAB} [A,B]=0\Longrightarrow
  L(A,B)=0.\end{equation}
Indeed,   let $\Gamma$ be a contour as in formula (\ref{eq:trQderiv}) along
a spectral ray around the spectrum of $A^{t_0}B$ for some fixed $t_0$, then
\begin{eqnarray*}
 \frac{d}{dt}_{\vert_{t=t_0}} \log(A^t B)
&=&\frac{i}{2 \pi} \int_\Gamma \log \lambda\,  \frac{d}{dt}_{\vert_{t=t_0}} (A^t
B-\lambda)^{-1}\, d\lambda\\
&=&\frac{i}{2 \pi} \int_\Gamma \log \lambda \, (A^{t_0}
B-\lambda)^{-1}\, \log A\,A^{t_0}
B\,   (A^{t_0}
B-\lambda)^{-1}\, d\lambda\\
&=&\log A\,A^{t_0}
B\, \frac{i}{2 \pi} \int_\Gamma \log \lambda    \,  (A^{t_0}
B-\lambda)^{-2}\, d\lambda \quad {\rm since }\quad [A,B]=0\\
&=&-\log A\,A^{t_0}
B\, \frac{i}{2  \pi} \int_\Gamma  \lambda^{-1}     (A^{t_0}
B-\lambda)^{-1}\, d\lambda \quad {\rm by }\quad {\rm integration}\quad {\rm by
}\quad {\rm parts}\\
&=&-\log A\, \,   A^{t_0}
B\,  (A^{t_0}
B)^{-1}\\
&=&- \log A.
\end{eqnarray*}
Similarly, we have $ \frac{d}{dt}_{\vert_{t=t_0}} \log(A^t )=-\log
A$ so that finally  $
\frac{d}{dt}_{\vert_{t=t_0}}L(A^t,B)=\frac{d}{dt}_{\vert_{t=t_0}}
\log(A^t B)-\frac{d}{dt}_{\vert_{t=t_0}} \log(A^t )$ vanishes. It
follows that $L(A,B)= \int_0^1 \frac{d}{dt}_{\vert_{t=\tau}}L(A^t,
B)\,
d\tau=0.$\\
Since $L(A,B)$ vanishes when $A$ and $B $ commute, weighted determinants are
multiplicative on commuting operators. \endsquare\\\\\indent
Let us now turn to the multiplicative anomaly for
$\zeta$-determinants, relating it  to weighted traces of $L(A,B)$.
An admissible  operator $A\in \Cl(M, E)$ with spectral cut $\theta$
and positive order  has well defined $\zeta$-determinant:
$${\rm det}_{\zeta,\theta}(A):= e^{-\zeta_{A,\theta}^\prime(0)}= e^{{\rm
tr}^A_\theta(\log_\theta A)}$$
since $\zeta_{A, \theta}(z):= {\rm TR}(A_\theta^{-z})$ is
holomorphic at $z=0$. In the second equality,  the weighted trace
has been extended to logarithms as before, picking out the constant
term of the meromorphic  map $z\mapsto {\rm TR}(\log_\theta  A\,
Q^{-z})$ (which can have double poles) with the notations of
section 2.  \\
 Recall from
\cite{PS} that
\begin{equation}\label{eq:detzeta}
\log {\rm det}_{\zeta, \theta}(A)= \int_M dx\, \left[{\rm TR}_x(\log_\theta A)-
\frac{1}{2a} {\rm
  res}_{x}(\log_\theta ^2A)\right]
\end{equation}
where $a$ is the order of $A$ and where ${\rm res}_{x}$ is the
noncommutative residue density extended to log-polyhomogeneous operators
defined previously. This expression corresponds to minus the coefficient in
$z$ of the Laurent expansion of TR$(A^{-z})$. \vskip 0,3cm
The $\zeta$-determinant generally depends on the choice of spectral
cut. However, it is invariant under mild changes of spectral cut in
the following sense.
\begin{lem}\label{lem:detspectralcut} Let $0\leq\theta<\phi<2\pi$ be two spectral
cuts for
the  admissible operator $A$.  If  there is a cone
$\Lambda_{\theta, \phi}$ (see \ref{eq:cone}) which does not
intersect the spectrum of the leading symbol of $A$ then
$${\rm det}_{\zeta, \theta}(A)={\rm det}_{\zeta, \phi}(A).$$
\end{lem}
{\bf Proof:} By (\ref{eq:detzeta}), and since $\log_\phi
A-\log_\theta A=2i\pi\Pi_{\theta, \phi}(A)$ is a finite rank operator and
hence smoothing under the assumptions of the proposition, we have\\
\begin{eqnarray*} \frac{{\rm det}_{\zeta,\phi}(A)}{{\rm
det}_{\zeta,\theta}(A)}& =& e^{\int_M dx\, \left[{\rm
TR}_x(\log_\phi A)- \frac{1}{2a} {\rm
  res}_{x}(\log_\phi^2A)\right]-\int_M dx\, \left[{\rm TR}_x(\log_\theta
A)- \frac{1}{2a} {\rm
  res}_{x}(\log_\theta^2A)\right]}\\
&=&e^{\int_M dx\, \left[{\rm TR}_x(\log_\phi A-\log_\theta A) -
\frac{1}{2a} {\rm
  res}_{x}(\log_\phi^2A-\log_\theta^2A)\right]}\\
&=&e^{\int_M dx\, \left[{\rm TR}_x(2i\pi\, \Pi_{\theta, \phi}(A)) -
\frac{1}{2a} {\rm
  res}_{x}\left((\log_\phi A+\log_\theta A)\, 2i\pi \, \Pi_{\theta,
\phi}(A)\right)\right]}\\
  &=& e^{2i\pi\,{\rm tr}( \Pi_{\theta, \phi}(A)) -
\frac{2i\pi}{2a} {\rm   res}\left((\log_\phi A+\log_\theta A)\,
\Pi_{\theta,\phi}(A)\right)}\\
  &=& e^{2i\pi \, {\rm rk} (\Pi_{\theta,\phi}(A))} \\
  &=& 1,
\end{eqnarray*}
where we have used the fact that the noncommutative residue vanishes on
smoothing operators on which the canonical trace coincides with the
usual trace on smoothing operators.
\endsquare\vskip 0,3cm\indent The $ \zeta$-determinant is not multiplicative \footnote{It was
shown in
  \cite{LP} that all multiplicative   determinants on
  elliptic operators can be built from two basic types of determinants; they
  do not include the $\zeta$-determinant.}.  Indeed, let   $A$ and $B$  be two
 admissible  operators with positive order and spectral cuts $\theta$
and $\phi$ and such that $AB$  is also admissible with spectral cut
$\psi$.  The multiplicative anomaly
$${\cal M}_\zeta^{\theta, \phi, \psi}(A, B):= \frac{{\rm det}_{\zeta,
\psi}(AB)}{{\rm det}_{\zeta, \theta}(A)\,{\rm
    det}_{\zeta, \phi}(B)},$$
 was proved to be local, independently by  Okikiolu \cite{O2} for operators
with scalar leading symbol and by Kontsevich and Vishik \cite{KV}
for operators ``close to identity'' (see  the introduction for a  more detailed
historical account).\\
 For simplicity, we drop the
explicit mention of $\theta, \phi, \psi$ and write
${\cal M}_\zeta(A, B).$ \vskip 0,3cm
Even though the operator $\log_\theta^2 A$ is not classical we have
the following useful property.
\begin{lem}\label{lem:logsquare}
 Let $A, B$ be admissible  operators in $\Cl(M, E)$ with
  positive orders $a, b$ and spectral cuts $\theta$ and $\phi$ respectively
and such that $AB$
(which is elliptic) is also admissible with spectral cut $\psi$. Then
$$K(A, B):=\frac{1}{2(a+b)} \log_\psi^2 A\,  B -\frac{1}{2a} \log_\theta^2A-
  \frac{1}{2b} \log_\phi^2B$$
has a symbol of the form
$$\sigma_K\sim {\rm ln} \vert \xi\vert (\sigma_0^{AB} -\sigma_0^A-\sigma_0^B)
+ \sigma_0^K$$ for some zero order classical symbol $\sigma_0^K$ and
where we have written $\sigma_{\log A}(x,\xi)=a\,  {\rm ln}\vert
\xi\vert I+\sigma_0^A (x,\xi)$ for an admissible operator $A$ of
order $a$.\\
In particular, both operators
$L(A,B)\, \frac{\log A}{a}-K(A,B)$ and $L(A,B)\, \frac{\log B}{b}-K(A,B)$ are
classical operators of zero order.
\end{lem}
 {\bf Proof:} By formula (\ref{eq:logspectralcut}), another choice of
spectral cut only changes the logarithms by adding an operator in
$\Cl^0(M, E)$ so that it will not affect the statement. As usual, we
drop the explicit mention of spectral cut assuming the operators
have common spectral cuts.
\\
An explicit computation on  symbols shows the result. Indeed, since
$\sigma_{\log A}(x, \xi)\sim a\, {\rm ln}\vert\xi\vert +\sigma_0^A(x, \xi)$,
we have\\
\begin{eqnarray*}
\sigma_{\log^2A}(x,\xi)&=& \sigma_{\log A}\star \sigma_{\log A}(x,\xi) \\
&\sim & a^2\, {\rm ln}^2 \vert\xi\vert I+2a\, {\rm
ln}\vert\xi\vert\,\sigma_0^A(x,\xi)
      +\sigma_0^A (x, \xi)\cdot \sigma_0^A(x,\xi)\\
&{}&+\sum_{\alpha\neq 0}{(-i)^{\vert
\alpha\vert}\over{\alpha!}}\partial^{\alpha}_{\xi}
      \sigma_0^A(x,\xi)\, \partial^{\alpha}_x\sigma_0^A(x, \xi).
\end{eqnarray*}
This yields:
\begin{eqnarray*}
\sigma_{K}(x,\xi)
&\sim &\ {\rm ln} \vert\xi\vert\left (\sigma_0^{AB}-\sigma_0^A-\sigma_0^B\right)
(x,\xi)\\
&{} & +\frac{1}{2(a+b)}\sigma_0^{AB}(x,\xi)\sigma_0^{AB}(x,\xi)+\sum_{\alpha\ne0}^{}{1\over{\alpha!}}\partial^{\alpha}_{\xi}
      \sigma_0^{AB}(x,\xi) D^{\alpha}_x\sigma_0^{AB}(x,
      \xi)\\
&{}&-\frac{1}{2a}\sigma_0^A(x,\xi)\sigma_0^A(x,\xi)-\sum_{\alpha\ne0}^{}{1\over{\alpha!}}\partial^{\alpha}_{\xi}
      \sigma_0^A(x,\xi) D^{\alpha}_x\sigma_0^A(x, \xi)\\
&{}&-\frac{1}{2b}\sigma_0^B(x,\xi)\sigma_0^B(x,\xi)-\sum_{\alpha\ne
    0}^{}{1\over{\alpha!}}\partial^{\alpha}_{\xi}
      \sigma_0^B(x,\xi) D^{\alpha}_x\sigma_0^B(x, \xi)
\end{eqnarray*}
from which  the first part of the statement follows.\\ On the other hand, it
follows from (\ref{eq:sigmaLAB}) combined with (\ref{eq:symbollog}) that the operators
$L(A,B)\, \frac{\log A}{a}$ and $L(A,B)\, \frac{\log B}{b}$ both have symbols which differ from ${\rm ln} \vert\xi\vert\left (\sigma_0^{AB}-\sigma_0^A-\sigma_0^B\right)
(x,\xi)$ by a classical symbol of order zero, from which we infer the second part of
the statement.
\endsquare \vskip 0,3cm \indent
The following theorem provides a local formula for the
multiplicative anomaly independently of  Okikiolu's assumption
that the leading symbols be scalar.
 \begin{thm}\label{thm:zetamultanom}  Let
$A$ and  $B$ be two admissible operators in
$\Cl(M, E)$ with
  positive orders $a, b$  and with spectral cuts $\theta$ and $\phi$  in $[0,2 \pi[$ such that
there is a cone delimited by the rays $L_\theta$ and $L_\phi$
which does not intersect the spectra of the leading symbols of $A$, $B$ and
$AB$. Then the product
$AB$  is admissible with a spectral cut $\psi$ inside that cone and
 the  multiplicative anomaly
${\cal M}_\zeta^{\theta, \phi, \psi}(A, B)$ is local    as a  noncommutative residue, independently of the
choices of $\theta, \phi,$ and $\psi$ satisfying the above requirements.\\
 Explicitly, and dropping the explicit mention of the spectral cuts, there is a classical  operator $W(\tau)(A,B)$ given by
(\ref{eq:Wtau}) of order
zero  depending continuously on $\tau$ such that:
 \begin{eqnarray}\label{eq:resmultiplicativeanomaly}
&{}&\log {\cal M}_\zeta(A, B)\nonumber\\
&=& \int_0^1{\rm
  res}\left( W(\tau)(A,B)\,\left(\frac{\log(A^\tau B)}{a\tau+b}
-\frac{\log B}{b}\right)\right) d\tau\nonumber\\
&+ & {\rm res} \left(
\frac{L(A, B)\,  \log B  }{b} - \frac{\log^2 A\, B}{2(a+b)}+\frac{\log^2A}{2a}+ \frac{\log^2B}{2b} \right)\nonumber\\
&=& \int_0^1{\rm
  res}\left( W(\tau)(A,B)\,\left(\frac{\log(A^\tau B)}{a\tau+b}
-\frac{\log A}{a}\right)\right) d\tau\nonumber\\
&+ & {\rm res} \left( \frac{L(A, B)\,  \log A }{a} - \frac{\log^2
A\, B}{2(a+b)}+\frac{\log^2A}{2a}+ \frac{\log^2B}{2b} \right)
\end{eqnarray}
When $A$ and $B$ commute the multiplicative anomaly reduces to:
\begin{eqnarray}\label{eq:resmultiplicativeanomalycomm}
\log {\cal M}_\zeta(A, B) &=& - {\rm res}\left( \frac{1}{2(a+b)}
\log ^2(A\,  B) -\frac{1}{2a} \log ^2A-
  \frac{1}{2b} \log ^2B\right)\nonumber\\
&=& \frac{ab}{2(a+b)} { \rm res} \left[ \left( \frac{\log
A}{a}-\frac{\log B}{b}\right)^2\right].
\end{eqnarray}
\end{thm}
\begin{rk}{\rm 
 For commuting operators, (\ref{eq:resmultiplicativeanomalycomm}) gives back
 the results of Wodzicki  as well as formula (III.3) in \cite{D}:
 $$\log {\cal M}_\zeta(A, B)= \frac{{\rm res}\left(\log^2(A^b B^{-a} )\right)}{2ab
(a+b)}.$$}
\end{rk}
{\bf Proof:}  As in the proof of the locality of the multiplicative
anomaly for weighted determinants (see Proposition
\ref{prop:weightedmultanom}), the independence of the choice of
spectral cuts satisfying the requirements of the  theorem follows
from Lemma \ref{lem:detspectralcut}.
\\ Combining equations (\ref{eq:detzeta}), the defect formula (\ref{eq:trQPS})
applied to the operator $L(A,B)$ and  weight $B$ with equation (\ref{eq:trQL})
applied to $Q=B$  we  write:
\begin{eqnarray}\label{eq:proof}
&{}&\log {\cal M}_\zeta(A, B)\nonumber\\
&=&
\log {\rm det}_\zeta(AB)- \log {\rm det}_\zeta(A)-\log {\rm det}_\zeta(B)\nonumber\\
&=&  \int_M dx\, \left[{\rm TR}_x(L(A,B)) \right.\nonumber\\
&{}&\left. -\left( \frac{1}{2(a+b)} {\rm
  res}_x(\log ^2AB)- \frac{1}{2a} {\rm
  res}_x(\log ^2A)- \frac{1}{2b} {\rm
  res}_x(\log ^2B)\right)\right]\nonumber\\
&=&{\rm tr}^B(L(A,B))+  \int_M dx\, \left[
\frac{1}{b}\, {\rm res}_x\left(L(A, B)\,  \log B  \right)
\right.\nonumber\\
&{}&\left. -\left( \frac{1}{2(a+b)} {\rm
  res}_x\left(\log^2 A\, B \right)-\frac{1}{2a} {\rm
  res}_x(\log ^2A)- \frac{1}{2b} {\rm
  res}_x(\log ^2B)\right)\right]\\
&=& \int_0^1{\rm
  res}\left( W(\tau)(A,B)\,\left(\frac{\log(A^\tau B)}{a\tau+b}
-\frac{\log B}{b}\right)\right) d\tau\nonumber\\
&+ & {\rm res} \left( \frac{L(A, B)\,  \log B  }{b} - \frac{\log^2
A\, B}{2(a+b)}+\frac{\log ^2A}{2a}+ \frac{\log ^2B}{2b}
\right),\nonumber
\end{eqnarray}
which proves the first equality  in (\ref{eq:resmultiplicativeanomaly}). The
second one can be derived similarly exchanging the roles of $A$ and $B$. \\
When $A$ and $B$ commute, by (\ref{eq:commutingLAB}), the operator $L(A,B)$ vanishes so that  (\ref{eq:proof}) reduces to:
\begin{eqnarray*}
\log {\cal M}_\zeta(A, B)
&=&{\rm tr}^B(L(A,B))+  \int_M dx\, \left[
\frac{1}{b}\, {\rm res}_x\left(L(A, B)\,  \log B  \right)
\right.\\
&{}&\left. -\left( \frac{1}{2(a+b)} {\rm
  res}_x\left(\log^2 A\, B \right)-\frac{1}{2a} {\rm
  res}_x(\log ^2A)- \frac{1}{2b} {\rm
  res}_x(\log ^2B)\right)\right]\\
&=&   -{\rm res}\left( \frac{\log^2 A\, B}{2(a+b)}-\frac{\log ^2A}{2a}- \frac{\log ^2B}{2b} \right)\\
&=&\frac{ab}{2(a+b)} { \rm res} \left[ \left( \frac{\log
A}{a}-\frac{\log B}{b}\right)^2\right].
\end{eqnarray*}
\endsquare
\vfill\eject
\noindent \bibliographystyle{plain}

\begin{thebibliography}{99}
\bibitem[B]{B}  N. Bourbaki, {\bf  El\'ements de math\'ematique: Alg\`ebre}
  Livre II,  Hermann, 1947
\bibitem[BGV]{BGV} N. Berline, E. Getzler, M. Vergne, {\bf Heat kernels and
    Dirac operators}, Springer Verlag  1992
\bibitem[BG]{BG} J.L. Brylinski, E. Getzler.
  {\it The homology of algebras of pseudodifferential symbols and non
    commutative residues}. $K$-theory, {\bf 1}, 385--403, 1987
\bibitem[CDMP]{CDMP}  A. Cardona, C. Ducourtioux, J.-P. Magnot, S. Paycha,
   {\it Infinite dimensional analysis, quantum probability and related
    topics}, Vol 5, no. 4 (2002) 503--40

\bibitem[D]{D} C. Ducourtioux, {\it Weighted traces on
    pseudodifferential operators and associated determinants}, PhD thesis,
  Clermont-Ferrand (2001)

\bibitem[F]{F} D. Freed,  {\it The geometry of loop groups},
  J. Diff. Geom. {\bf 28} (1988) 223-276


\bibitem[Fr]{Fr}  L. Friedlander, PhD Thesis, Dept. Math. MIT 1989

\bibitem[FrG]{FrG}  L. Friedlander, V. Guillemin, {\it Determinants of zeroth
    order operators} math.SP/0601743
\bibitem[G]{G} V. Guillemin, {\it   A new proof of
Weyl's formula on the asymptotic distribution of eigenvalues}, Adv.
Math. {\bf 55} (1985) 131--160

\bibitem[H]{H} S. Hawking, {\it Zeta function regularization of path integrals
    in curved spacetime}, Comm. Math. Phys. {\bf 55} (1977) 133-148

\bibitem[KV]{KV} M. Kontsevich, S. Vishik,
{\it Geometry of determinants of elliptic operators}, Func.
Anal. on the Eve of the XXI century, Vol I, Progress in Mathematics
{\bf 131} (1994)  173--197 ; \otherterm{ Determinants of elliptic
pseudodifferential operators}, Max Planck Preprint (1994)

\bibitem[K]{K} Ch. Kassel,\otherterm{ Le r\'esidu non commutatif [d'apr\`es
    Wodzicki]}, S\'em. Bourbaki {\bf 708} (1989)

\bibitem[L]{L} M. Lesch, \otherterm{
On the non commutative residue for pseudodifferential operators
with log-polyhomogeneous symbols}, Ann.  Global  Anal.
Geom. {\bf 17} (1998)  151--187

\bibitem[LP]{LP} J.-M. Lescure, S.Paycha, {\it Traces on pseudodifferential
 operators and associated determinants} (2005)  Proc. Lond. Math. Soc. (3)
{\bf 94} no. 3 (2007) 772--812

\bibitem[MN]{MN} R. Melrose, V. Nistor, {\it
Homology of pseudodifferential operators I.
Manifolds with boundary}, funct-an/9606005, june
1999
\bibitem[O1]{O1}
K. Okikiolu, \otherterm{The
   Campbell-Hausdorff theorem for elliptic operators and a related
  trace formula}, Duke. Math. Journ.  {\bf 79}  (1995) 687--722

\bibitem[O2]{O2} K. Okikiolu, \otherterm{The multiplicative anomaly for
    determinants of elliptic oprators}, Duke Math. Journ. {\bf 79} (1995)
722--749
\bibitem[P]{P} S. Paycha, {\it Renormalised traces as a looking glass into
    infinite-dimensional geometry}, Infin. Dimens. Anal. Quantum
  Probab. Relat. Top.  {\bf 4}, no. 2 (2001)
221--266
\bibitem[Po1]{Po1} R. Ponge, {\it Spectral asymmetry,
    zeta functions and the noncommutative residue},
    internat. J. Math. {\bf 17} no. 9 (2006) 1065--1090
\bibitem[Po2]{Po2} R. Ponge,{\it Traces on pseudodifferential operators and
    sums of commutators},  To appear in J. Anal. Math..
\bibitem[PS]{PS} S. Paycha, S. Scott, {\it A  Laurent expansion for
    regularised integrals of holomorphic symbols}, Geom.  Funct. Anal. {\bf 17} no.
2 (2007) 491--536
 \bibitem[RS]{RS} D.B. Ray, I.M. Singer, {\it
    R-torsion and the Laplacian on Riemannian manifolds}, Adv. Math.
  {\bf t7}  (1971) 145--210
\bibitem[Sc]{Sc} S. Scott, {\it The  residue determinant}, Commun. Part. Diff. Eqn.s
{\bf 30} no. 4-6
 (2005) 483--507

\bibitem[Se]{Se} R.T. Seeley, {\it Complex
powers of an elliptic operator, Singular integrals}, Proc. Symp. Pure Math.,
Chicago, Amer. Math. Soc.,
 Providence  (1966) 288--307

\bibitem[W1]{W1} M. Wodzicki, {\it Non commutative residue. Chapter I. Fundamentals}
in Lecture
  Notes in Math. {\bf 1289}  320-399, Springer
Verlag 1987; {\it Spectral asymmetry and  noncommutative residue}
(in Russian) Thesis, (former) Steklov Institute, Sov. Acad. Sci.
Moscow 1984
\bibitem[W2]{W2} M. Wodzicki, Commentary, in {\it Hermann Weyl's selected papers}
  (in Russian), edited by V.I. Arnold and A.N. Parshin, Nauka Moscow, 1985
\end{thebibliography}

\vskip 10mm {\small \noi\textsc{Laboratoire de Math\'ematiques,
Complexe des C\'ezeaux, Universit\'e Blaise Pascal, 63 177
Aubi\`ere Cedex F. E-mail:} sylvie.paycha@math.univ-bpclermont.fr

\vskip 5mm \noi \textsc{D\'epartment de Math\'ematiques,
Universit\'e de Ouagadougou, 03 bp 7021.
  Burkina Faso. E-mail}
marie.oued@univ-ouaga.bf}
\end{document}